\documentclass[sigconf]{acmart}

\usepackage{algorithm}
\usepackage{algorithmic}
\usepackage{bm} 
\usepackage{color}
\usepackage{colortbl}
\usepackage{float}
\usepackage{lscape}
\usepackage{makecell}
\usepackage{multirow}
\usepackage{subcaption}
\usepackage{soul}

\newcommand{\proposed}{PrISM-Observer}

\newcommand{\tabref}[1]{Table~\ref{#1}}
\newcommand{\figref}[1]{Figure~\ref{#1}}
\newcommand{\secref}[1]{Section~\ref{#1}}

\newcommand{\appref}[1]{Appendix~\ref{#1}}

\newcolumntype{C}[1]{>{\centering\arraybackslash}m{#1}}

\newcommand{\etal}{\textit{et al.}}
\newcommand{\eg}{\textit{e.g.},~}
\newcommand{\ie}{\textit{i.e.},~}

\AtBeginDocument{%
  \providecommand\BibTeX{{%
    \normalfont B\kern-0.5em{\scshape i\kern-0.25em b}\kern-0.8em\TeX}}}

\setcopyright{acmcopyright}
\copyrightyear{2024}
\acmYear{2024}
\setcopyright{rightsretained}
\acmConference[UIST '24]{The 37th Annual ACM Symposium on User Interface Software and Technology}{October 13--16, 2024}{Pittsburgh, PA, USA}
\acmBooktitle{The 37th Annual ACM Symposium on User Interface Software and Technology (UIST '24), October 13--16, 2024, Pittsburgh, PA, USA}
\acmDOI{10.1145/3654777.3676350}
\acmISBN{979-8-4007-0628-8/24/10}

\begin{document}

\title[\proposed{}]{\proposed{}: Intervention Agent to Help Users Perform Everyday Procedures Sensed using a Smartwatch}

\author{Riku Arakawa}
\orcid{0000-0001-7868-4754}
\affiliation{%
  \institution{Carnegie Mellon University}
  \city{Pittsburgh}
  \country{USA}
}
\email{rarakawa@cs.cmu.edu}

\author{Hiromu Yakura}
\orcid{0000-0002-2558-735X}
\affiliation{
    \institution{Max-Planck Institute for Human Development}
    \city{Berlin}
    \country{Germany}
}
\email{yakura@mpib-berlin.mpg.de}

\author{Mayank Goel}
\orcid{0000-0003-1237-7545}
\affiliation{%
  \institution{Carnegie Mellon University}
  \city{Pittsburgh}
  \country{USA}
}
\email{mayankgoel@cmu.edu}

\renewcommand{\shortauthors}{Arakawa et al.}

\begin{teaserfigure}
    \centering
    \includegraphics[width=\linewidth]{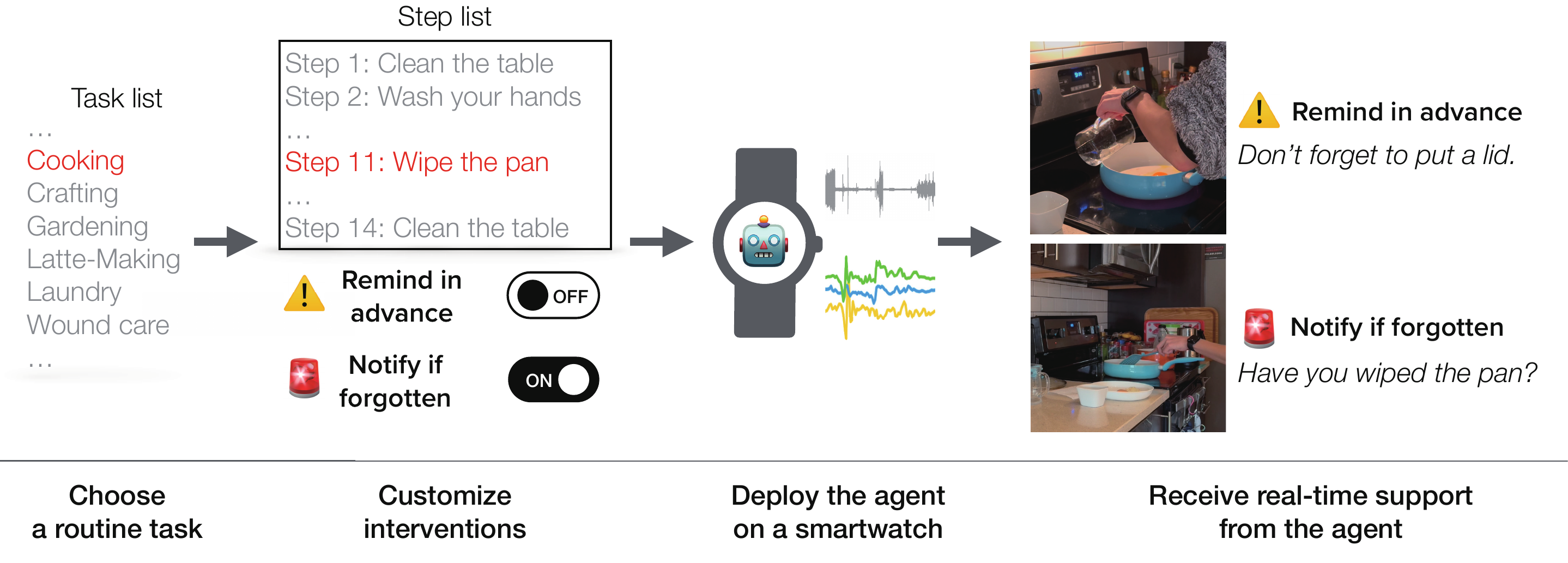}
    \caption{\proposed{} is a framework to design and trigger interventions to mitigate user errors in daily procedural tasks such as cooking. The user actions are sensed using a smartwatch, which can be used to deliver the interventions and integrate into the user's life to observe their task execution. The system combines multimodal sensing (sound + motion) with stochastic modeling of user actions to forecast the intervention moment. It reminds users of key steps at the optimal moment and notifies them if they forget a step in real-time.}
    \Description{This image illustrates a four-step process for using a smart technology aid in routine tasks. The first step displays icons representing routine tasks such as drinking coffee, cooking, and washing hands. The second step shows a list of steps for a task, such as 'Step 1: Clean the table' and 'Step 11: Wipe the pan,' with toggles for advance reminders and notifications for forgotten actions. The third step depicts a smartwatch deploying an intelligent agent, represented by a brain icon with signal waves. The final step shows photographs of a person being reminded by the smartwatch with alerts like 'Have you wiped the pan?' and 'Don't forget to clean the wand,' indicating the proposed interaction in assisting with the daily tasks.}
    \label{fig:teaser}
\end{teaserfigure}

\begin{abstract} 
We routinely perform procedures (such as cooking) that include a set of atomic steps.
Often, inadvertent omission or misordering of a single step can lead to serious consequences, especially for those experiencing cognitive challenges such as dementia.
This paper introduces \textit{\proposed{}}, a smartwatch-based, context-aware, real-time intervention system designed to support daily tasks by preventing errors.
Unlike traditional systems that require users to seek out information, the agent observes user actions and intervenes proactively.
This capability is enabled by the agent's ability to continuously update its belief in the user's behavior in real-time through multimodal sensing and forecast optimal intervention moments and methods.
We first validated the steps-tracking performance of our framework through evaluations across three datasets with different complexities.
Then, we implemented a real-time agent system using a smartwatch and conducted a user study in a cooking task scenario.
The system generated helpful interventions, and we gained positive feedback from the participants.
The general applicability of \proposed{} to daily tasks promises broad applications, for instance, including support for users requiring more involved interventions, such as people with dementia or post-surgical patients.
\end{abstract}



\maketitle

\section{Introduction}
\label{sec:intro}

Every day, we perform many tasks, ranging from cooking to crafting to self-care, which involve a series of atomic steps.
Accurately executing all the steps of tasks can be difficult, especially when the tasks become routine and fail to capture our full attention~\cite{Hu2023Effect} or when we face cognitive challenges such as dementia~\cite{Lancioni2021Technology}.
For example, people often forget to turn on the washing machine after loading it or turn off lights before leaving home~\cite{MemoryLapse}.
Such mistakes where we omit essential steps or confuse the order of actions can lead to undesirable outcomes~\cite{reason1990human}.
For instance, in a study with over a hundred participants, close to 20\% of participants made critical errors while using COVID-19 self-test kits~\cite{Pydi2023}.
Thus, real-time assistance by sensing a user's actions and intervening as needed can help improve quality of life.

Most of the existing task-support solutions in HCI are tailored for specific activities, often necessitating specialized equipment or advanced computers equipped with cameras and displays, such as Augmented Reality (AR) glasses.
For instance, Uriu~\etal~\cite{DBLP:conf/chi/UriuNTKIO12} developed a sensor-equipped frying pan that offers context-sensitive information like the pan's current temperature.
Also, \textit{AdapTutAR}~\cite{DBLP:conf/chi/HuangQWPSCRQ21} is a machine task tutoring system designed to monitor users in following tutorials and offer feedback through AR glasses.
However, we often need help with mundane tasks or in situations where instrumenting ourselves or the environment is not easy. 
There is a need for a solution that seamlessly integrates into supporting various routine tasks and is practical for constant use.
Additionally, current systems predominantly rely on users actively seeking information, like consulting a recipe while cooking~\cite{DBLP:conf/tei/SatoWR14, DBLP:conf/chi/UriuNTKIO12}.
There has been limited exploration into passive interactions, where systems proactively monitor and offer corrective feedback when errors occur.
Designing such interactions is complex, as the system needs to model the user's spontaneous behavior to predict errors and intervene without becoming intrusive or annoying.
The challenge intensifies when we shift focus from camera-based methods to more practical and ubiquitous methods such as motion and sound sensors on smartwatches. 
These solutions offer seamless assimilation into daily life while lowering privacy concerns but come at the cost of sensor accuracies.
Our prior work~\cite{DBLP:journals/imwut/ArakawaYMNRDRMC22} tried addressing the errors of sound and motion-based Human Activity Recognition (HAR) models by combining procedure knowledge with multimodal sensor data.
However, we have yet to use these advances to build a reliable intervention system that guides a user through a variety of tasks in real-time.

To achieve an agent system that intervenes to mitigate errors in daily procedural tasks, this paper introduces \textit{\proposed}. 
It preemptively models user task behavior and optimizes intervention methods and timing by considering uncertainties for the sensing data and anticipating the user's future behavior.
Moreover, the framework allows the design of user-friendly reminder-based interventions that the user or a system designer can customize.
Using the framework, we developed a prototype system that operates on a smartwatch (\figref{fig:teaser}) -- a device chosen for its ubiquity, minimal privacy concerns compared to camera-based systems, and capability to monitor a user across various daily activities.
The prototype offers timely and relevant interventions with minimal reliance on task-specific rules.

\proposed{} either reminds users to execute a step in advance (\textsc{remind in advance}), or if it infers that the user may have forgotten a step, it notifies them separately (\textsc{notify if forgotten}). 
These interventions are time-critical as a reminder that comes too early or too late will be useless.
To verify \proposed{}'s ability to optimize the intervention timing, we applied the proposed framework to three tasks with different procedural complexities: wound care, cooking, and latte-making (Study~1). 
The results showed the proposed approach reduced the timing error compared to a baseline approach that does not use sensor information: by averaging across all steps, $26.8$ (baseline) $\rightarrow 24.1$ (proposed) seconds in the wound care, $119.5 \rightarrow 61.5$ seconds in the cooking, and $50.1 \rightarrow 22.3$ seconds in the latte-making tasks, respectively.

Subsequently, we built a real-time agent system that assisted the users in a specific cooking task of making a sunny-side-up and a grilled sausage.
We evaluated the agent through a user study (Study~2, $N=10$) to examine the system's performance and usability.
The results showed that the participants perceived the triggered interventions to be accurate (20 out of the total 27 triggered interventions were scored five or higher for accuracy on the 7-point Likert scale in the post-task questionnaire).
We also found a trend between the delay and perceived accuracy for the two kinds of intervention, highlighting the importance of optimizing their timing.
In addition, the \textsc{notify if forgotten} interventions were triggered correctly (22 out of the total 25 intervention chances), verifying the intervention policy's effectiveness.
Overall, the participants found the system reliable and showed a positive behavioral intention (8 out of 10 participants). 
Furthermore, their comments validated our design of using different types of interventions and making them customizable, as well as offered implications for the interaction between humans and real-world task-support agents.

In this paper, we make the following contributions:
\begin{enumerate}
    \item framework for modeling user behavior in procedural tasks and designing interventions, building upon the foundation of the existing multimodal procedure tracking module~\cite{DBLP:journals/imwut/ArakawaYMNRDRMC22}.
    \item comprehensive evaluation across three daily-task datasets demonstrated the proposed approach's superiority in forecasting the optimal moments for interventions, highlighting its effectiveness in complex tasks with over 50\% timing error reduction.
    \item user study with a real-time prototype system on a smartwatch, which not only showed its preferable experience but also informed design implications to build reliable and acceptable task-support agents.
\end{enumerate}
It is important to note that \proposed{} performs promisingly, even though the underlying HAR models are approximately only 50\% accurate at detecting each atomic step of the procedures. 
This result demonstrates that it is possible to use imperfect sensing and machine learning to build a useful intervention system to aid an imperfect human prone to making mistakes.
Our system will be particularly helpful for users who face cognitive challenges (\eg patients with dementia or after surgery), and we are working closely with such populations as part of our future work.
We open-source the framework and the dataset to facilitate the research in this domain (\url{https://github.com/cmusmashlab/prism}).

\section{Background and Related Work}
\label{sec:rw}

We first examine the cognitive psychology literature on human errors in daily tasks, recognizing that such errors are inevitable and necessitate support.
Next, we review HCI studies concerning assistants for procedural tasks, underscoring the importance of using a prevalent device to support everyday procedures and its unique challenges.
Finally, we explore multimodal sensing research related to activity recognition, especially using a smartwatch.

\subsection{Human Errors in Everyday Tasks}
\label{sec:rw-error}

The improper execution of everyday tasks, ranging from cooking and medical self-care to machine use, is a multifaceted issue rooted in cognitive psychology.
Studies have shown that errors are likely to occur when the working memory load is high or the user is not fully attentive~\cite{Gray2000Milliseconds, Byrne1997Working}.
For instance, multitasking during cooking can lead to oversights or mistakes in recipe execution~\cite{Hu2023Effect}.
Also, Beaver~\etal~\cite{Beaver2017Characterising} discussed that complex and integrative steps in daily activities may be the first to be affected by cognitive decline.
In fact, cognitive challenges such as memory lapses contribute significantly to non-adherence to self-care activities by the elderly~\cite{Park1992Medication}.
Additionally, misunderstanding instructions can easily lead to critical errors; almost 20\% of participants made mistakes while using COVID-19 self-test kits~\cite{Pydi2023}. 
These examples underscore the human error-prone nature in various daily scenes~\cite{reason1990human}, highlighting the need for situated support that helps users avoid or recover from errors to compensate for cognitive limitations, as emphasized by Zhang~\cite{Zhang2023Underlying}

\subsection{Assistants for Procedural Tasks}
\label{sec:rw-assistant}

Supporting users in conducting complicated tasks in the real world has been a popular research theme in HCI research.
A common approach involves crafting specialized devices tailored for particular activities or tasks.
For example, Lee and Dey~\cite{DBLP:conf/chi/LeeD14} developed a sensor-augmented pillbox and feedback system to improve medication compliance.
In the cooking domain, \textit{Cooking Navi}~\cite{DBLP:conf/mm/HamadaOISST05} is an interface providing multimedia recipe information (\ie text, video, and audio) to aid in cooking.
Uriu~\etal~\cite{DBLP:conf/chi/UriuNTKIO12} extended this support system by creating a sensor-equipped frying pan that offers context-sensitive information like the pan's current temperature.
Similarly, \textit{MimiCook}~\cite{DBLP:conf/tei/SatoWR14} combines a depth camera and projector to deliver on-the-spot guidance during cooking.

Computer-vision-based approaches are popular to guide users through various tasks~\cite{Servn2012Assembly, DBLP:conf/hci/HasadaZYRO19, DBLP:conf/uist/YamaguchiMMT0SK20, DBLP:conf/chi/HuangQWPSCRQ21, DBLP:conf/chi/LiuZJHVQ0R23}.
For instance, \textit{AR Cooking}~\cite{DBLP:conf/hci/HasadaZYRO19} used 3D animation of cookware on AR glasses.
Serván~\etal~\cite{Servn2012Assembly} developed a system to overlay work instruction in an assembly task using AR.
Similarly, \textit{AdapTutAR}~\cite{DBLP:conf/chi/HuangQWPSCRQ21} is a machine task tutoring system that monitors learners' tutorial-following status and provides feedback via AR glasses.
\textit{HoloAssist}~\cite{DBLP:conf/iccv/WangKRPCABFTFJP23} is a system where a human observer watches the task performer’s egocentric video captured by AR glasses and guides them verbally.
To support the creation of such technologies, researchers have compiled datasets capturing first-person perspectives on procedural tasks such as cooking~\cite{DBLP:journals/corr/abs-2312-14556} or assembly~\cite{DBLP:conf/cvpr/SenerCSHSWY22, schoonbeek2024industreal}. 

Despite the success of these systems, using cameras and displays can result in privacy-invasive and power-hungry systems. 
This issue becomes apparent when we want to support users' daily routines pervasively in contrast to specific, high-stakes situations such as assembly~\cite{Riedel2021Deep}.
Furthermore, these solutions predominantly offer context-aware information, assuming that users actively seek the information.
While benefiting those inexperienced with the task, these systems may not assist adept users who might still commit errors due to inattentiveness or cognitive overload.
A similar motivation for error-checking systems was discussed by Bovo~\etal~\cite{DBLP:conf/iui/BovoBBJ20}, who proposed an approach for real-time error prediction in sequence-constrained procedural tasks.
While showing promising performance in detecting errors in an item-picking-placement task, the method relies on the assumption of a predefined step sequence and task-specific heuristics like item location, which does not apply to various tasks where users are permitted to exhibit multiple behavioral patterns.

Hence, this paper aims to create a generalizable framework to monitor and intervene with users engaged in everyday tasks.
Given such circumstances, users might not always prefer to utilize sophisticated equipment like AR glasses or external cameras for every task. 
Consequently, given its widespread use and minimal interference with the task, we have chosen a smartwatch as our device.
This decision undoubtedly introduces a research question: \textit{how can we design a reliable intervention agent system for users' spontaneous behavior using imperfect sensing?}
Our solution is to stochastically model user behavior from sensor observation with transition knowledge to trigger situated interventions.
We use minimized task-specific assumptions to offer flexibility and cater to different user preferences.

\subsection{Smartwatch-Based Context Sensing}
\label{sec:rw-smartwatch}

Human Activity Recognition (HAR) is a widely studied technology that senses user actions and behavior~\cite{DBLP:journals/kbs/YadavTPS21}.
Among many modalities used for HAR, smartwatch-based systems often use audio and motion data~\cite{DBLP:conf/uist/LaputXH16,DBLP:journals/imwut/GuanP17, DBLP:conf/uist/LaputAGH18, DBLP:conf/huc/BeckerFS19, DBLP:journals/imwut/MollynAVHG22}.
For example, \textit{ViBand}~\cite{DBLP:conf/uist/LaputXH16} enabled bio-acoustic sensing using commodity smartwatch accelerometers, detecting the use of different hand-held tools such as a toothbrush and heat gun.
Ashry~\etal~\cite{Ashry2020CHARM} used cascading bidirectional long short-term memory to classify motion data of different daily activities.
For audio, \textit{Ubicoustics}~\cite{DBLP:conf/uist/LaputXH16} employed acoustic sensing to classify 30 daily activities, such as hand washing and typing.
Recent work has proposed a multimodal learning approach to maximize the capability, for instance, \textit{GestEar}~\cite{DBLP:conf/huc/BeckerFS19}, which combined audio and motion signals to distinguish different gestures, such as knocking and snapping.
Moreover, \textit{SAMoSA}~\cite{DBLP:journals/imwut/MollynAVHG22} explored the downsampling of the audio modality by supplementarily using the motion data to make the system more privacy-aware.
These advances in the smartwatch's HAR capability led to deployed applications, such as Apple Watch's Handwash detection feature~\cite{handwashing}.

We aim to create an intelligent agent that assists users in preventing errors in various tasks.
Here, despite its potential, a significant challenge with HAR is its limited accuracy when extended to a wide range of real-world activities. 
For instance, Liaqat~\etal~\cite{DBLP:conf/mobisys/LiaqatWGARL18a} highlighted the difficulties posed by noise in tasks performed outside controlled environments.
Additionally, when HAR is employed for tracking procedural tasks, distinguishing between steps with similar signal profiles is often hard.
In this regard, studies like those by Nakauchi~\etal~\cite{DBLP:conf/ro-man/NakauchiSTM09} and Arakawa~\etal~\cite{DBLP:journals/imwut/ArakawaYMNRDRMC22} have suggested the potential of enhancing tracking accuracy by using transition information between steps of a task.
While these approaches have improved the accuracy of procedure tracking, the performance is still far from perfect.
Thus, a fully functional interactive system for procedural task support on a smartwatch has yet to be realized.
This work provides an approach to using state-of-the-art activity recognition and procedure tracking algorithms to design a reliable agent system.

\section{Framework for Error Monitoring in Procedural Tasks}
\label{sec:proposed}

In this section, we introduce a framework for monitoring users' actions and offering timely interventions.
We first outline a user scenario to illustrate the utility of such an agent system and introduce our intervention design.
Then, we present an algorithm using multimodal sensing to forecast user actions and a policy to trigger interventions.

\subsection{User Scenario}
\label{sec:proposed-scenario}

As discussed in \secref{sec:rw-error}, a user's performance in daily tasks is affected by various factors, such as inattention and cognitive load or time pressure. 
Consider two scenarios: \textit{1) Tom decided to prepare a sunny-side-up for breakfast instead of his usual choice of scrambled egg. Given he is not used to this task, he inadvertently forgets to add oil to the pan before cracking an egg. Consequently, the egg sticks to the surface of the pan. When Tom attempts to lift it with a spatula, the yolk breaks, leading to a disappointing start to the day.}
\textit{2) Catherine is preparing a latte using a semi-automatic coffee machine in her office's shared kitchen. She is in a hurry to get to her next meeting and forgets to clean the steam wand after use. As a result, milk residue clogs the holes of the wand, necessitating maintenance.}
In these situations, it would be helpful if a system could remind users just in time to avoid errors or notify them when errors are detected in real-time.
We envision a context-aware agent living on the user's watch that works in the following way:
\textit{1) Just as Tom is about to crack the egg, he receives a reminder on the watch to pour oil into the pan.}
\textit{2) When Catherine is about to leave the kitchen with her latte, her watch notifies her to clean the steam wand.}
The following subsections describe the formulation and implementation to achieve such monitoring and intervention.

\subsection{Interaction Design}
\label{sec:proposed-formulation}

\begin{figure}[t]
    \centering
    \includegraphics[width=\linewidth]{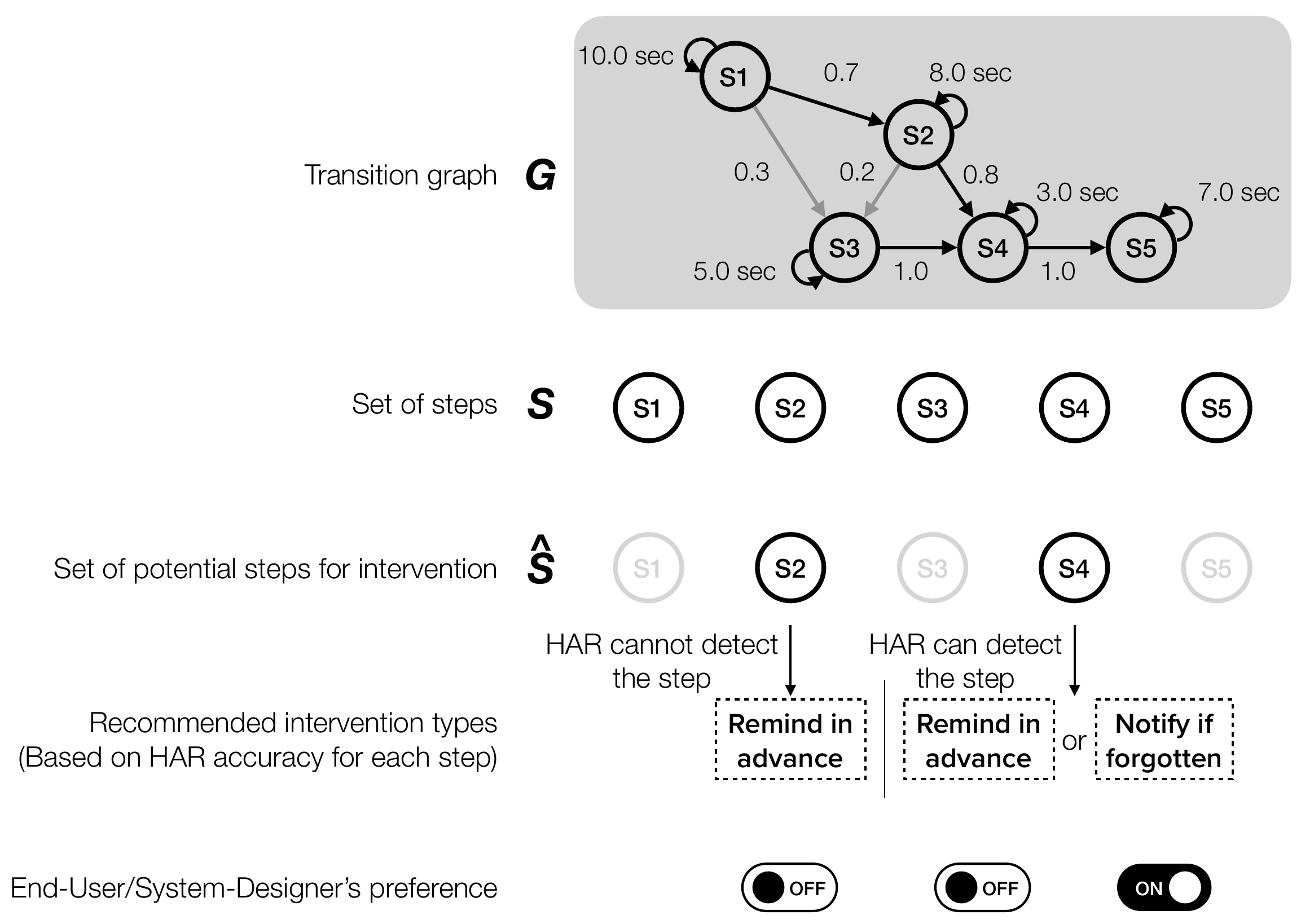}
    \caption{Process of selecting steps for intervention and their types. The system suggests the possibility of the \textsc{notify if forgotten} intervention based on the HAR accuracy of the step. Refer to~\secref{sec:proposed-formulation} for details.}
    \Description{This image presents a structured approach to step selection and intervention in a task flow. At the top, a transition graph G shows a sequence of steps S1 to S5, with arrows indicating the direction of progression and labeled with transition probabilities and time intervals. Below the graph, 'Set of potential steps for intervention S_hat' is shown with highlighted circles S2 and S4 as potential intervention points. Below, there are 'Recommended intervention types' correlating with human activity recognition (HAR) accuracy for each step. This section suggests intervention types, such as 'Remind in advance' for non-detectable steps and 'Notify if forgotten' for steps with higher accuracy in detection. The bottom row shows toggles representing an end-user or task-designer’s preference, with 'Remind in advance' off and 'Notify if forgotten' on, indicating customizable intervention preferences.}
    \label{fig:step-selection}
\end{figure}

To enable support for a diverse set of tasks, we design our interactions in a generalizable manner instead of using task-specific heuristics as discussed in~\secref{sec:rw-assistant}.
Let $S = \{s_1, s_2, \dots s_N\}$ be the set of the atomic steps in the procedure where $N$ is the total number of the steps.
We assume a transition graph $G$, which contains information on the average time spent on each step and the transition probabilities between steps.
$G$ can be obtained from sample demonstrations of the task as described by Arakawa~\etal~\cite{DBLP:journals/imwut/ArakawaYMNRDRMC22}.

Current research on procedure tracking treats each step of a process equally, focusing on accuracy in recognizing each step on a frame-by-frame basis~\cite{DBLP:conf/ro-man/NakauchiSTM09, DBLP:journals/imwut/ArakawaYMNRDRMC22}. 
In contrast, to develop a user-centered helpful agent, it is crucial to prioritize the steps each user needs support for. 
Considering Tom's cooking example above, it might not be necessary to verify whether he remembered to bring an egg from the refrigerator, as this action inevitably happens during the cooking process.
Therefore, we assume a subset $\hat{S}\in S$ as a set of steps for which situated interventions can be helpful.

Given the nature of each step in $\hat{S}$, suitable interventions may change.
For example, users may want to receive notifications only when they forget a specific step, \ie error detection. 
Or, users might appreciate receiving preemptive reminders for crucial steps, especially when the timing and sequence are critical and irreversible -- for instance, adding oil to the pan before cracking an egg into it.
Providing users with a global control like this is key to successfully developing human-AI interaction systems~\cite{DBLP:conf/chi/AmershiWVFNCSIB19}.
Accordingly, we prepared two types of interventions that will be assigned to each target step $\hat{s} \in \hat{S}$: \textbf{\textsc{remind in advance}} and \textbf{\textsc{notify if forgotten}}.
The \textsc{remind in advance} intervention is intended to happen before the user starts the step.
On the other hand, the \textsc{notify if forgotten} intervention only happens when the user forgets the step.
To reliably achieve this notification (\ie preventing false positives or negatives), the system must be able to detect whenever the step happens accurately so that, when the step is not inferred, the system can be confident that the user missed the step.
Thus, we enabled the system to suggest the possibility of using the \textsc{notify if forgotten} intervention for a step based on the step's detectability from sensors, the detail of which is described in~\secref{sec:proposed-policy}, hence satisfying a key requirement for successful human-AI interaction -- communicating how well the system functions~\cite{DBLP:conf/chi/AmershiWVFNCSIB19}.
In addition, \proposed{}'s default intervention messages support users' efficient dismissal in case of false predictions without annoying them~\cite{DBLP:conf/chi/AmershiWVFNCSIB19}.
For the \textsc{remind in advance} intervention, the message is ``Don't forget to do $\hat{s}$,'' while it is ``Have you done $\hat{s}$?'' for the \textsc{notify if forgotten} intervention.

Lastly, system designers or end-users can finalize the configuration, \ie which $\hat{s} \in \hat{S}$ are supported by interventions and the intervention type.
For instance, a kitchen manager in Catherine's office may be particularly interested in ensuring that cleaning occurs post-use of the machinery, thus enabling the \textsc{notify if forgotten} intervention for the step.
\figref{fig:step-selection} summarizes the entire selection process for intervention. 

\subsection{Intervention Timing Optimization}

In our framework, the system persistently observes user behavior to provide situated interventions, triggering them at opportune times.
A significant obstacle arises from the fact that the training data often lacks instances of actual user errors, and it is hard to train a model to predict or detect errors directly.
Thus, identifying omitted actions or reminding before a specific step in real-time is particularly challenging, especially when steps can be completed in various sequences, unlike prior work assuming a constrained sequence~\cite{DBLP:conf/iui/BovoBBJ20}.
To address this, we propose an alternative strategy to forecast when a specific step, $\hat{s}$, should occur based on the current belief about the user state -- preemptively estimating the remaining time till $\hat{s}$ happens.
We refer to this remaining time till the user reaches $\hat{s}$ from the current step (at time $t$) by using a stochastic variable $D_t^{\hat{s}}$, which we describe in detail later in the next subsection.
At a high level, this framework decides whether to trigger an intervention based on the expectation of when the user will perform the step $\hat{s}$.
Once the certainty of the timing of a step goes above an empirically determined threshold, \proposed{} prepares a \textit{timer} to trigger an intervention corresponding to the step.
This timer functions like this: if the system estimates the user will be doing a certain step in 10 seconds with high confidence, it waits for \textit{slightly less than} 10 seconds to trigger the \textsc{remind in advance} intervention, or the \textsc{notify if forgotten} if the user does not do the step around those \textit{slightly more than} 10 seconds. 

Algorithm~\ref{alg:framework} presents the overview of the framework.
In this pseudo-code, for simplicity, we assume a single target step $\hat{s}$, but the framework is extendable to multiple target steps in parallel, as demonstrated in the user study later.
In the following subsections, we discuss the modeling of $D_t^{\hat{s}}$ and intervention policy.

\begin{algorithm}[t]
\caption{\proposed{}'s algorithmic framework}
\begin{algorithmic}[1]
\STATE \textbf{User Input:} $\hat{s}$ and its intervention type
\STATE \textbf{System Configuration:} trained \textit{PrISM-Tracker}~\cite{DBLP:journals/imwut/ArakawaYMNRDRMC22} (including transition graph $G$), threshold $h^{\hat{s}}$, offset constants $K^{+}$ and $K^{-}$
\STATE begin the task
\WHILE{the user has not finished the task}
    \STATE get the latest frame sensor data
    \STATE apply PrISM-Tracker and update the internal state
    \STATE calculate $E[D_t^{\hat{s}}]$ and $H[D_t^{\hat{s}}]$ (\secref{sec:proposed-stochastic})
    \IF{intervention timer should begin (\secref{sec:proposed-policy})}
        \IF {intervention is \textsc{remind in advance}}
            \STATE timer $\leftarrow$  $E[D_t^{\hat{s}}] - K^{-}$ 
        \ELSIF {intervention is \textsc{notify if forgotten}}
            \STATE timer $\leftarrow$  $E[D_t^{\hat{s}}] + K^{+}$ 
        \ENDIF
        \STATE timer begins
    \ENDIF
    \IF{timer ends}
        \IF {intervention is \textsc{remind in advance}}
            \STATE trigger the intervention
        \ELSIF {intervention is \textsc{notify if forgotten} \textbf{and} $\hat{s}$ has not been detected by HAR}
            \STATE trigger the intervention
        \ENDIF
    \ENDIF
\ENDWHILE
\STATE end the task
\end{algorithmic}
\label{alg:framework}
\end{algorithm}

\subsubsection{Stochastic Modeling of User Behavior}
\label{sec:proposed-stochastic}

We use a stochastic variable $D_t^{\hat{s}}$, indicating a remaining time at the given $t$ till the user reaches the target step $\hat{s}$.
$D_t^{\hat{s}}$ follows a probabilistic distribution $P(D_t^{\hat{s}})$.
The expectation can be calculated as,
\begin{align*}
E[D_t^{\hat{s}}] &= \sum D_t^{\hat{s}} P(D_t^{\hat{s}}) \\
&= \sum_{s \in S} P(s) \sum_{\tau \in T_s^{\hat{s}}} P(\tau) \text{Time}(s \rightarrow \hat{s}, \tau)
\end{align*}
, where $T_s^{\hat{s}}$ is the set of the possible trajectories from step $s$ to the target step $\hat{s}$, $P(s)$ indicates the probability of the user being at step $s$ at time $t$, $P(\tau)$ indicates the probability of the user will follow the trajectory $\tau$, and $\text{Time}(s \rightarrow \hat{s}, \tau)$ means the average time it takes to transition from step $s$ to $\hat{s}$ by following the path $\tau$.
$P(s)$ is obtained by PrISM-Tracker~\cite{DBLP:journals/imwut/ArakawaYMNRDRMC22}'s output while $P(\tau)$ and $\text{Time}(s \rightarrow \hat{s}, \tau)$ are calculated based on the transition graph $G$.

Importantly, this equation tells us two uncertainties with respect to user behavior: uncertainty in the current state (first sigma term) and uncertainty in the future trajectory (second sigma term).
To gauge the total uncertainty the system has, entropy can be calculated as,
\begin{align*}
H[D_t^{\hat{s}}] &= -\sum P(D_t^{\hat{s}}) \log P(D_t^{\hat{s}})
\end{align*}
Higher entropy means more uncertainty, meaning a variance in the estimation of $D_t^{\hat{s}}$.
Conversely, a low entropy value indicates that the user behavior is more predictable.
To computationally calculate the expectation $E[D_t^{\hat{s}}]$ and entropy $H[D_t^{\hat{s}}]$ in real-time, we used the Monte Carlo method~\cite{metropolis1949monte} with the sample size of 10,000.
Additionally, $T_s^{\hat{s}}$ is enumerated with depth-first search over $G$ from the current step $s$ to the target step $\hat{s}$.

\begin{figure}[t]
    \centering
    \includegraphics[width=\linewidth]{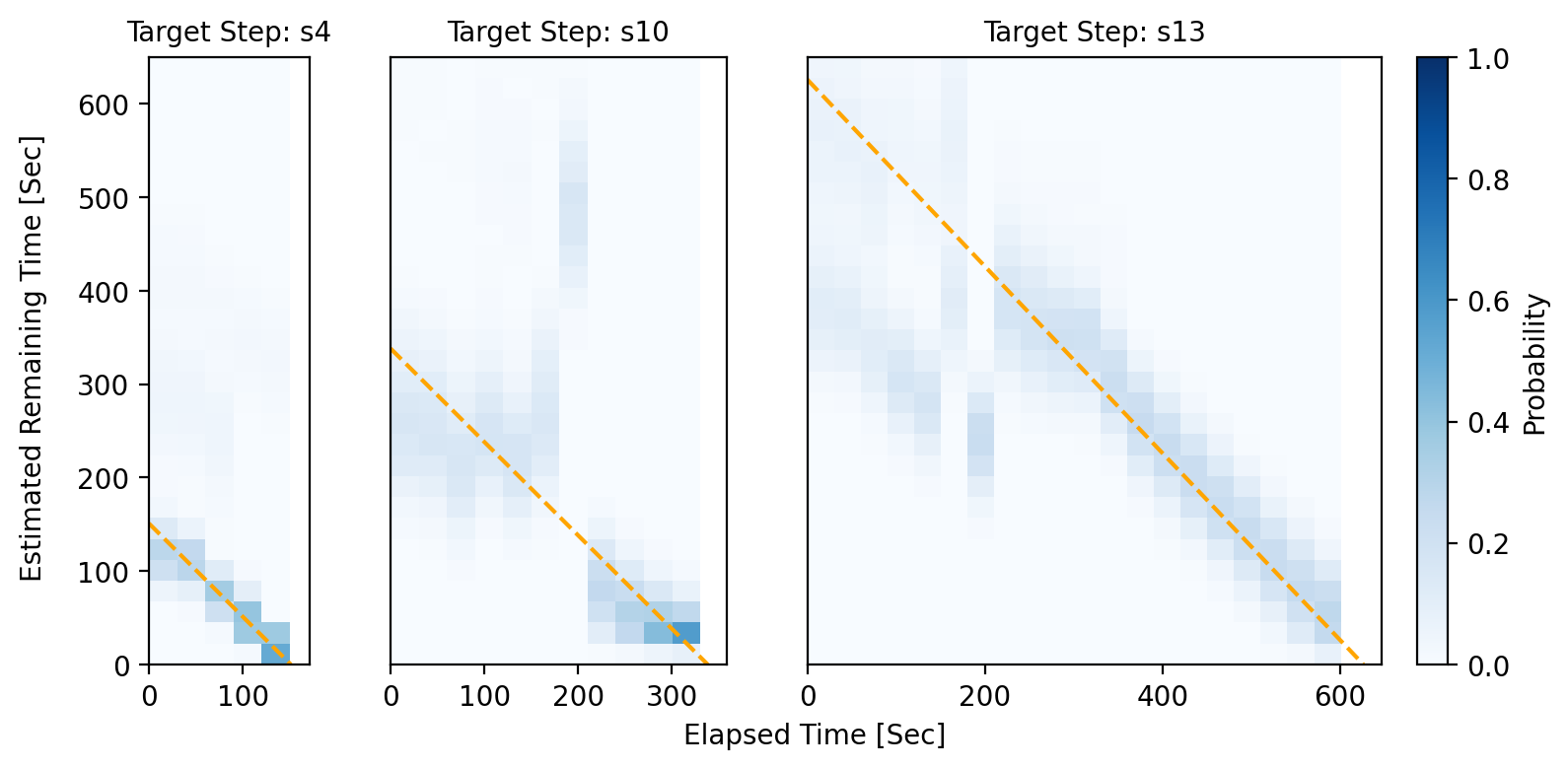}
    \caption{Example $D_t^{\hat{s}}$ transition during a session of the cooking task, where $\hat{s}$ is $s_4$ (performed early), $s_{10}$ (performed in the middle), and $s_{13}$ (performed later). The orange line is the ground truth (actual remaining time till the target step). Two uncertainties affect the distribution: current belief about the user state and future user trajectory.}
    \Description{This image contains three heatmaps representing the probability of the remaining time till a certain step in a task over time. Each heatmap corresponds to a target step, labeled s4, s10, and s13, respectively. The x-axis represents elapsed time in seconds, while the y-axis shows the estimated remaining time, also in seconds. Darker shades indicate a higher probability. The heatmaps feature a diagonal line, depicted with a dashed orange line. This line goes from the upper left corner to the lower right corner, suggesting a decrease in remaining time as more time elapses. The color bar on the right-hand side shows the probability scale from 0.0 to 1.0.}
    \label{fig:dt-distribution}
\end{figure}

\begin{figure*}[t]
    \centering
    \includegraphics[width=0.78\linewidth]{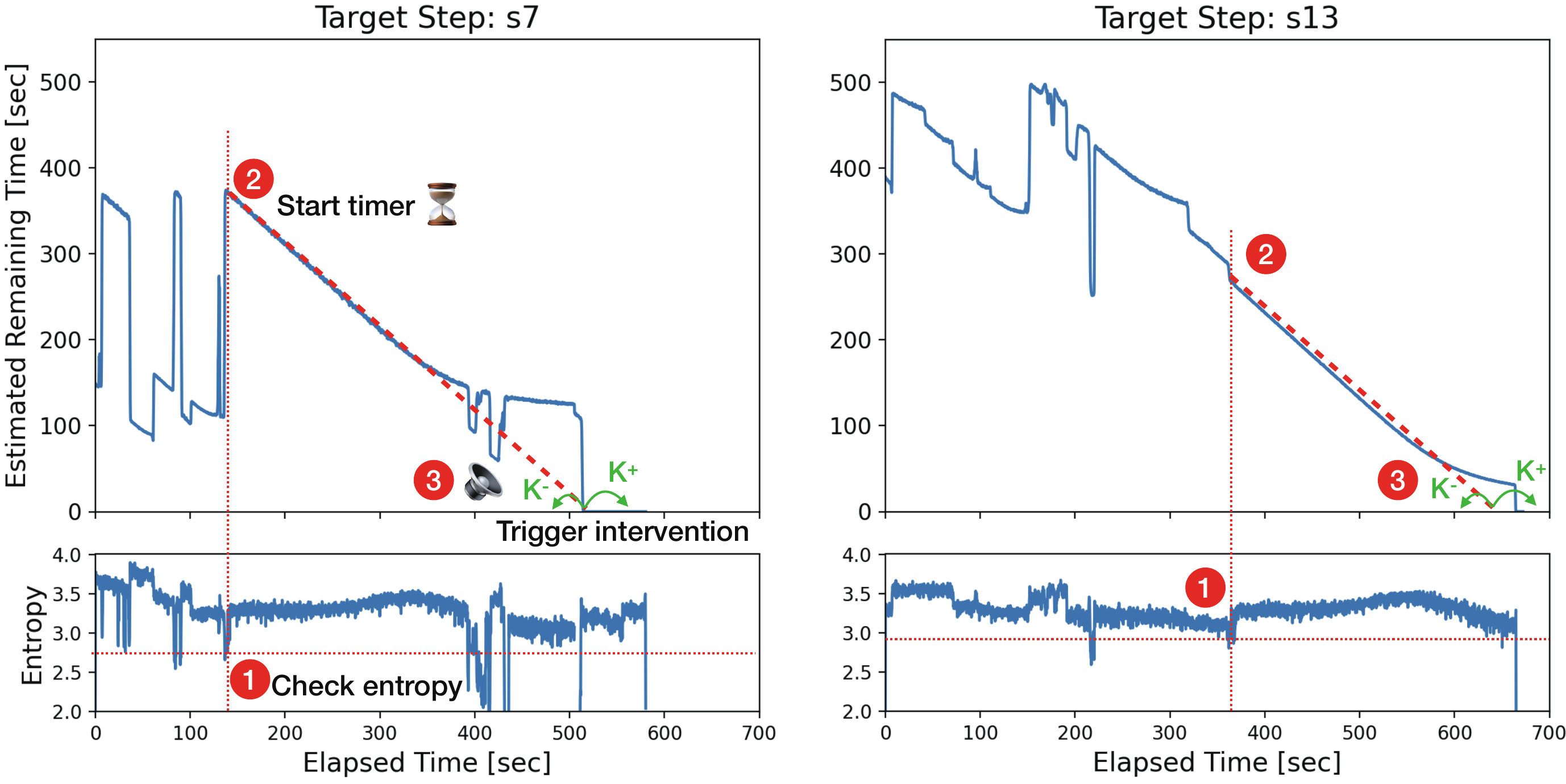}
    \caption{Example transitions of estimated remaining time $E[D_t^{\hat{s}}]$ and entropy $H[D_t^{\hat{s}}]$ from different sessions with different target steps and how our intervention policy works. The y-axis for the entropy graph starts from 2.0 for visualization. $K^-$ and $K^+$ are system parameters for the timing offsets of the \textsc{remind in advance} and \textsc{notify if forgotten} interventions, respectively.}
    \Description{This image features two sets of graphs, each representing transitions of estimated remaining time and entropy for different task sessions with distinct target steps, s7 and s13. The top graphs plot the estimated remaining time in seconds on the y-axis against elapsed time in seconds on the x-axis. They show a fluctuating line with an annotated three-step intervention policy: '2. Start timer' marked by a red dotted vertical line, and '3. Trigger intervention' marked by a bell icon and a green bracket indicating the timing offset parameters K- and K+. The bottom graphs represent entropy, with the y-axis starting at 2.0 for better visualization and the x-axis the same as above. The entropy levels fluctuate over time, with a notable red dotted line at '1. Check entropy,' indicating when the system triggers the timer ('2. Start timer').}
    \label{fig:timer}
\end{figure*}

Example distribution of $D_t^{\hat{s}}$ calculated from one session data (cooking task consisting of 14 steps, which we detail in \secref{sec:study1-dataset}) is shown in \figref{fig:dt-distribution}.
In general, the distribution gets more certain over time till the target step because the uncertainties are mitigated by sensor observation.
For instance, the estimation gradually becomes highly certain when the target step is $s_4$ (\figref{fig:dt-distribution}~Left).
On the other hand, when the target step is $s_{10}$ or $s_{13}$ (\figref{fig:dt-distribution}~Center\&Right), it has a high variance in the beginning primarily because there are several possible paths to the step (majorly whether cooking an egg or a sausage first).
After approximately 200 seconds, the variance diminishes when the system detects the step of pouring an egg on the pan, resolving the uncertainty by considering the sensor observation and transition graph.

\subsubsection{Intervention Policy}
\label{sec:proposed-policy}

Lastly, we describe how the intervention is triggered based on the distribution of $E[D_t^{\hat{s}}]$. 
The key idea is to preemptively forecast the timing of the target step and prepare the intervention.
To achieve this, the system continuously characterizes user behavior by $E[D_t^{\hat{s}}]$ and $H[D_t^{\hat{s}}]$.
\figref{fig:timer} shows the example transitions related to $s_7$ and $s_{13}$ in different sessions of the cooking task.
The expected remaining time (in \figref{fig:timer}~Top) gets reduced as the user performs each step and gets closer to the target step.
The measured entropy can remain quite noisy (as demonstrated in \figref{fig:timer}~Bottom-Left) due to uncertainties in sensor data and future user behavior. 
However, we look for brief moments of certainty for the model where it has a reasonable idea of when the target step might happen. 

The system starts a timer once the entropy ($H[D_t^{\hat{s}}]$) gets lower than a step-dependent hyperparameter $h^{\hat{s}}$, and $E[D_t^{\hat{s}}]$ is stable for a certain period afterward (``1'' in \figref{fig:timer}).
We provide a further detailed implementation of this timer policy in~\appref{app:policy-implementation-detail}.
This approach prevents the model from being confused later by sensor uncertainty, especially when the user takes unexpected behavior near the target step.
Of course, this policy can still lead to imperfections, but we will quantify its effectiveness for different tasks in Study~1.

Once the timer starts (``2'' in \figref{fig:timer}), the duration of the timer depends on the type of intervention. 
In case of the \textsc{remind in advance} intervention, the timer is set for $K^{-}$ seconds before the anticipated moment for the step of interest.
At the end of the timer, the notification is triggered (``3'' in \figref{fig:timer}) to remind the user to make sure they do the step.
In case of \textsc{notify if forgotten} intervention, the timer is set for ${K^{+}}$ \textit{after} the anticipated moment for the step of interest. 
Until the timer runs out, the system waits for the user to do the step of interest. 
If the user does not perform the step, the system generates a \textsc{notify if forgotten} intervention.
The values of constants $K^-$ and $K^+$ are decided by a system designer or can be adjusted by the end-user.
For example, if a user wants a notification as soon as possible after they skip a step, a smaller $K^{+}$ is chosen, but it might lead to noisy interactions. 
We envision designers will experiment with optimal values for these constants to tailor the interventions to their goals. 

\section{Study~1: Algorithm Evaluation in Multiple Daily Tasks}

We first investigate the effectiveness of the proposed algorithm in preemptively predicting the timing of specific steps.
We use a dataset of multiple daily procedures with different task complexity.

\subsection{Dataset}
\label{sec:study1-dataset}

\begin{figure*}[t]
    \centering
    \includegraphics[width=\linewidth]{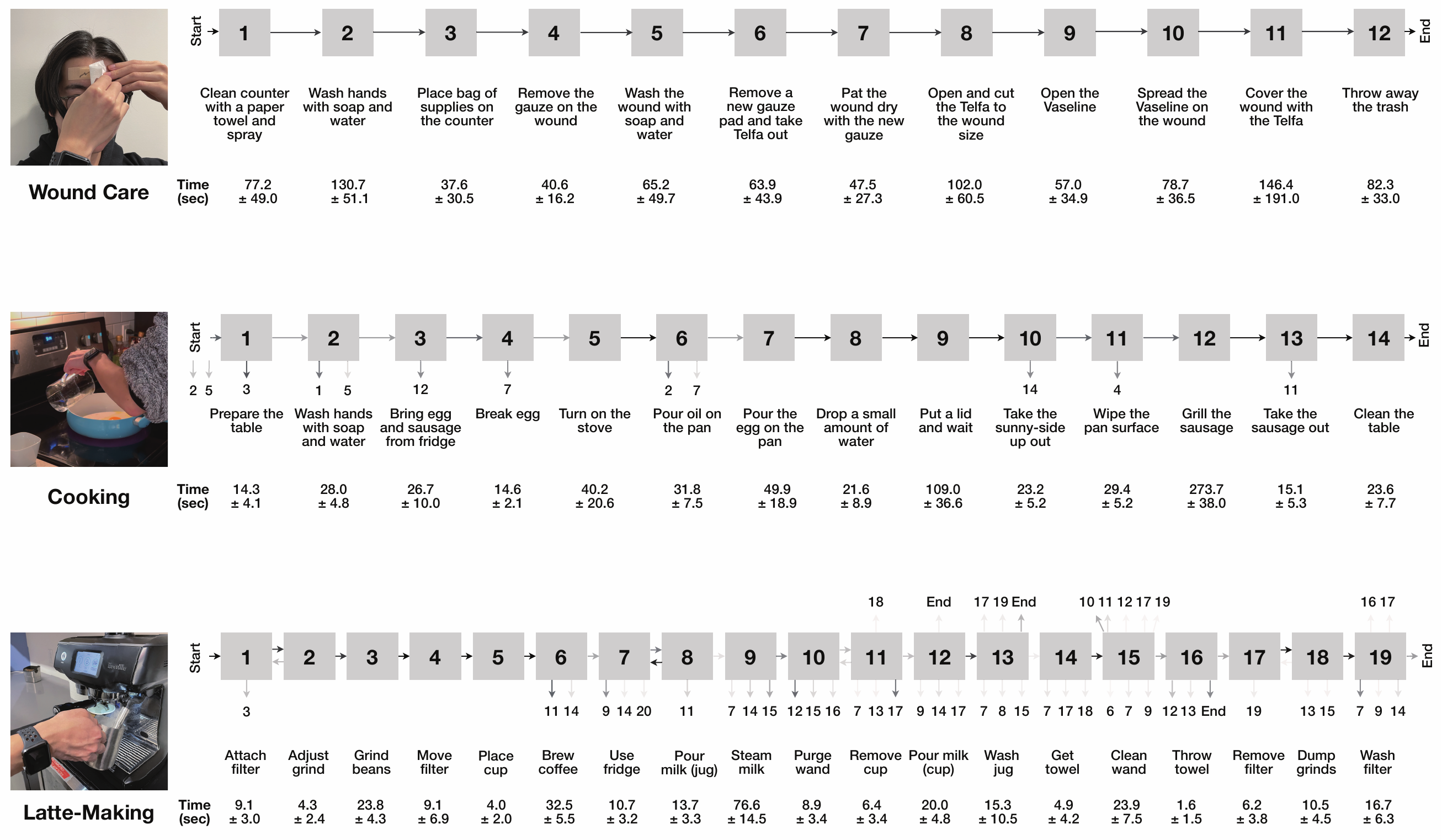}
    \caption{Transition graph of the three tasks: wound care, cooking, and latte-making. Each task has a different level of graph complexity, \ie the number of branches. Each arrow points to a possible transition step, and the opacity of the arrow represents the probability of the transition.}
    \Description{This image contains three process flow charts depicting routine tasks - Wound Care, Cooking, and Latte-Making, each accompanied by a series of photographs illustrating each step. The 'Wound Care' graph has 12 steps starting with cleaning the counter and ending with throwing away the trash. For example, Step 4 is 'Remove the gauze on the wound' with an average time of 40.6 seconds. The photograph shows a person applying medical gauze to an arm. The 'Cooking' chart details a 14-step process for preparing a meal, starting with preparing the table and finishing with cleaning the table. Step 7, 'Turn on the stove,' has an average time of 40.2 seconds. The corresponding image depicts a person interacting with a stovetop. Lastly, the 'Latte-Making' chart outlines a 19-step procedure for making a latte, beginning with attaching a filter and concluding with washing the grinder. Step 11 is 'Pour milk (cup)' with an average time of 20.0 seconds, with the photo illustrating pouring milk into a cup. Each step in the charts has a number associated with it, arrows to the next step, and the average time taken for the step, with a standard deviation indicated below.}
    \label{fig:dataset-graph}
\end{figure*}

We used the dataset of procedural tasks introduced in~\cite{DBLP:journals/imwut/ArakawaYMNRDRMC22}, that is, \textit{latte-making} (22 sessions, 15 participants) and \textit{wound care} (23 sessions, 23 participants) tasks.
Following the same protocol, we expanded the dataset by collecting new data on a \textit{cooking} task with 17 sessions and 8 participants who were already familiar with the task.
This data collection occurred in a single kitchen.
Note that, unlike the latte-making and wound care tasks where the watch was worn on the right wrist, the watch was placed on the left for the cooking task.
This difference in the setting led to differences in the motion data patterns since the participants were all right-handed, and thus, the watch might not have captured some crucial motions performed by the dominant hand.
Still, we anticipated that this new cooking scenario -- featuring users with the watch on their non-dominant hand's wrist -- would enhance the system's usability and reflect more natural use cases.
We preprocessed and obtained frame-level classification results using PrISM-Tracker~\cite{DBLP:journals/imwut/ArakawaYMNRDRMC22} with the same frame length of 0.2 seconds, to which we applied our framework.

These three tasks encompass a range of procedural complexity, distinguished by the number of branching paths within each task.
Wound care is a training task during perioperative counseling for skin cancer patients in a medical facility.
Given the clinical staff trains the patients on how to clean their post-surgical wounds, the task is linear, where the sequence of steps is predetermined.
However, given the individual variance in dexterity of patients, there is a high variance in the duration of each step across participants.
In contrast, the latte-making task is done by regular users of a coffee machine in an academic building.
These users were not given set instructions on how to use the machine.
Thus, the task involves more flexibility, allowing for various sequences of steps to be performed with minimal restrictions on ordering.
The cooking task represents an intermediate level of complexity.
It includes 14 steps where users prepare a sunny-side-up and grilled sausage.
The participants can choose the order in which they cook each item, which creates a major branch in the transition.
\figref{fig:dataset-graph} summarizes the transition graphs for the three tasks.
Notably, the dataset did not include error cases, and all steps were properly executed.
We will test our system's real-time capability to detect errors in Study~2.

\subsection{Metric}

\begin{figure}[t]
    \centering
    \includegraphics[width=0.8\linewidth]{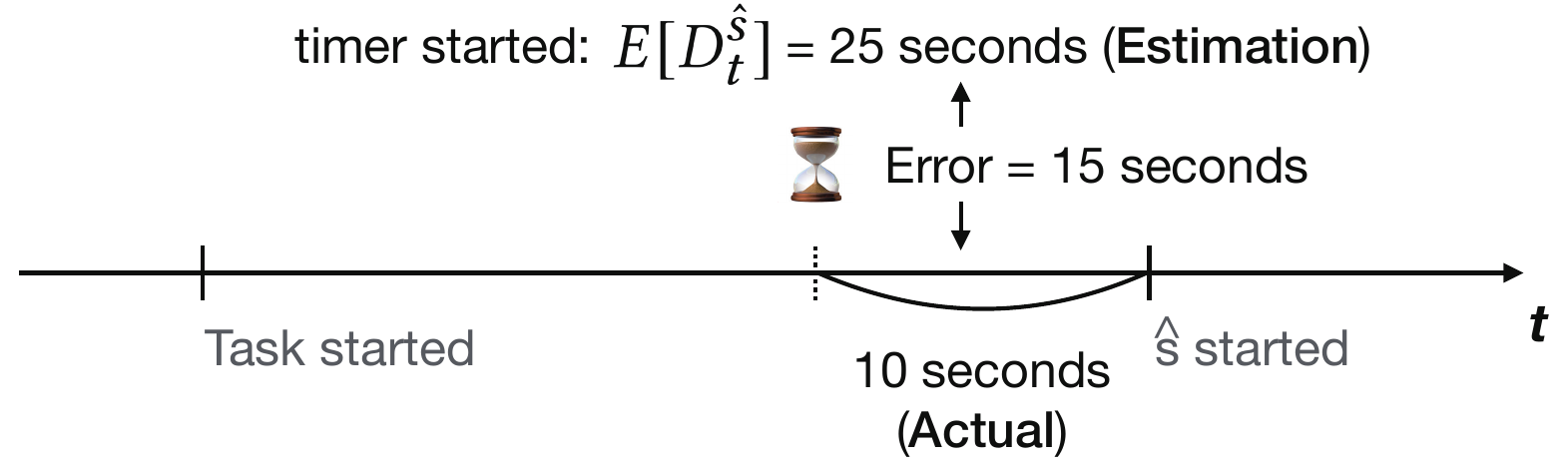}
    \caption{Example error calculation in the forecasted and actual timing of the target step in Study~1.}
    \Description{This image displays a timeline to illustrate the error in estimating a step’s start time. The horizontal line represents the time axis 't' with a marked point labeled 'Task started.' A point above the time axis indicates when a timer was started, labeled with 'E= 25 seconds (Estimation),' accompanied by a clock icon. A bidirectional arrow connects this point to another on the time axis marked '10 seconds (Actual),' indicating an actual start time of the task. The distance between the estimated and actual start times is noted with 'Error = 15 seconds,' showing the discrepancy between estimation and ground truth.}
    \label{fig:study1-metric}
\end{figure}

Study~1 aims to measure the algorithm's accuracy in forecasting the timing of future target steps.
To evaluate this, we processed session data from the beginning to emulate real-time prediction.
When the intervention timer started at time $t$ according to a policy, we calculated the error between the expected remaining time $E[D_t^{\hat{s}}]$ and the actual remaining time till the target step.
The example of this error calculation is presented in~\figref{fig:study1-metric}.
Also, while the system will be designed for certain target steps ($\hat{S}$) as discussed in~\secref{sec:proposed-formulation}, we assumed $\hat{S} = S$ to examine the accuracy for all steps in this study.

\subsection{Compared Intervention Policies}

We used the policy described in~\secref{sec:proposed-policy} as the \textit{proposed} condition.
We conducted leave-one-session-out cross-validation, and hyperparameters $h^{s_i} (s_i \in S)$ in this policy were obtained by grid search in each training fold.

In addition, we prepared the \textit{baseline} policy, which is based on the expected time to the target step at the beginning, \ie $E[D_0^{\hat{s}}]$. 
This policy serves as a condition where the system does not use the sensor data and relies on the transition history to predict the timing of the target step.

\subsection{Results}

\begin{figure}[t]
    \centering
    \includegraphics[width=\linewidth]{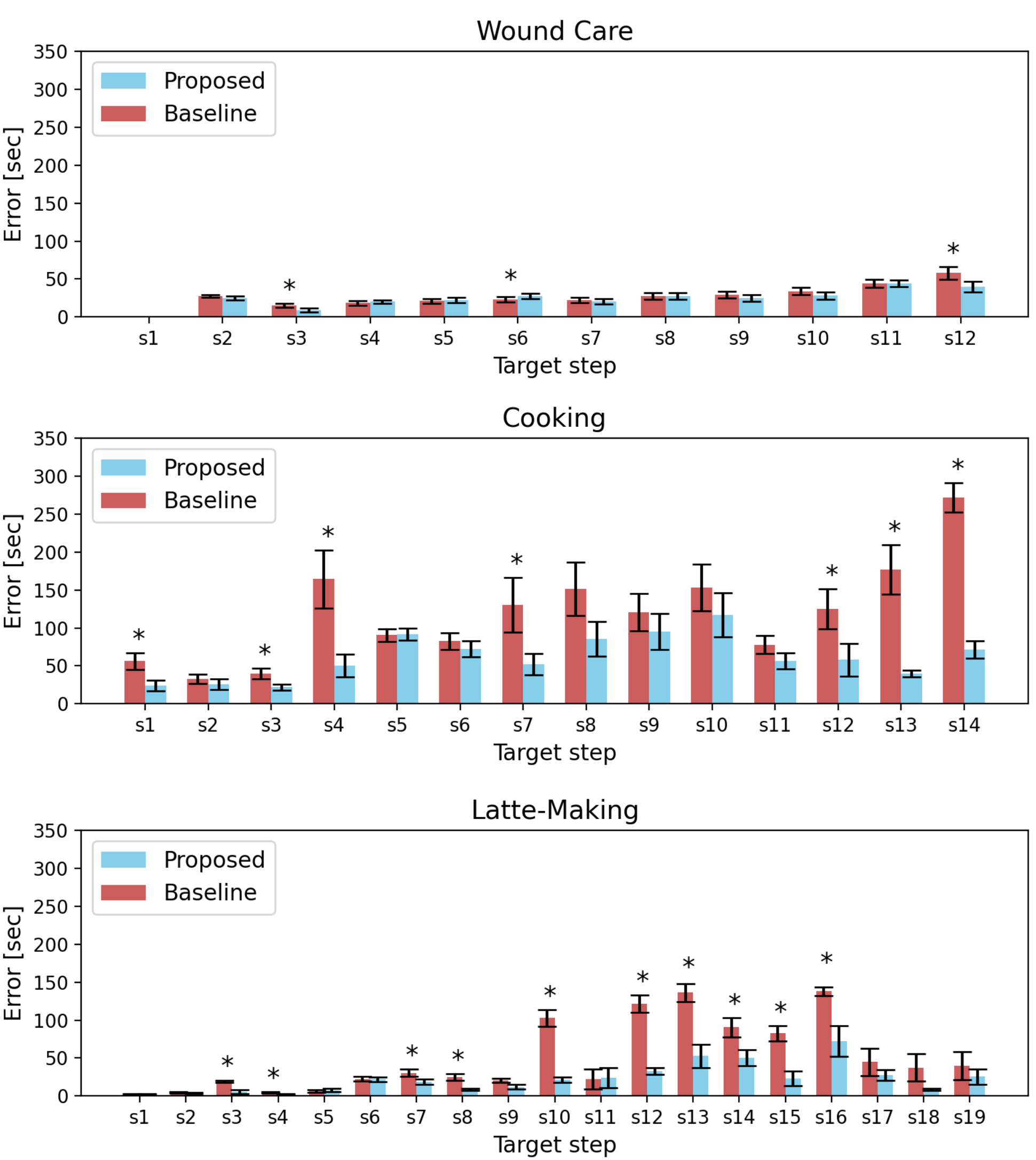}
    \caption{Result of Study~1 comparing the \textit{baseline} and \textit{proposed} policies in three daily procedural tasks. The \textit{proposed} policy greatly reduces the timing error as the procedure's complexity increases thanks to the sensor observation and the introduced stochastic modeling of user behavior. Error bars indicate standard error. * indicates $p<0.05$ in a two-sided paired t-test.}
    \Description{This image comprises three bar charts comparing error rates between a proposed method and a baseline across different tasks—Wound Care, Cooking, and Latte-Making. In each chart, the x-axis lists target steps (from s1 to s12 for Wound Care, s1 to s14 for Cooking, and s1 to s19 for Latte-Making), while the y-axis measures error in seconds. Two sets of bars are presented for each step: one in blue for the 'Proposed' method and one in red for 'Baseline.' The bars show the mean error, and the black lines represent the standard error. Asterisks are placed to highlight significant differences in the proposed and baseline approaches. Overall, the charts visually convey that the proposed method generally results in lower error rates than the baseline.}
    \label{fig:study1-result}
\end{figure}

The absolute timing error in the three tasks by different target steps $\hat{s}$ is presented in~\figref{fig:study1-result}.
Overall, the \textit{proposed} policy reduced the timing error from the \textit{baseline} policy at varying target steps, especially in tasks with more complexity.
At the same time, we observe the \textit{proposed} policy did not contribute much to reducing the errors at certain steps.
We delve deeper into the result of each task and discuss the effect of task complexity and sensing reliability on the error.

\subsubsection{Wound Care}

Both policies led to small errors (26.8 seconds for the \textit{baseline} policy and 24.1 seconds for the \textit{proposed} policy on average across all steps).
The satisfactory performance of the \textit{baseline} policy can be attributed to the single-threaded nature of the task.
Furthermore, the tracker's accuracy for this task is limited, as the steps do not introduce significant differences in sensor readings, particularly for intermediate steps ($s_3, ..., s_9$), as illustrated in~\figref{fig:wound-care-raw-result} in~\appref{app:prism-tracker-performance}.
The confusion led to the increased error for $s_6$ in the \textit{proposed} policy.
The result suggests that, when the sensing consists of much uncertainty, the system struggles to model user behavior from the observation.
Still, when the sensing accurately detects some actions, it benefits the \textit{proposed} policy.
For instance, the tracker's effective detection of state $s_2$ and $s_{11}$ significantly lowers the error in predicting the subsequent state $s_3$ and $s_{12}$ in the \textit{proposed} policy.

\subsubsection{Cooking}

The \textit{proposed} policy largely reduced the timing error compared to the \textit{baseline} policy (119.5 seconds for the \textit{baseline} policy and 61.5 seconds for the \textit{proposed} policy).
In the \textit{baseline} policy, the lack of context as to which item to cook first led to a large error, while the \textit{proposed} policy could infer it from the observation.
Similarly to the wound care result, it is implied that the larger errors in the \textit{proposed} policy (\ie $s_8, s_9, s_{10}$) came from the low sensing accuracy of $s_7$ and $s_8$, as shown in~\figref{fig:cooking-raw-result} in~\appref{app:prism-tracker-performance}, in addition to the relatively longer duration of $s_9$.

\subsubsection{Latte-Making}

Lastly, the \textit{proposed} policy again largely reduced the timing error compared to the \textit{baseline} policy (50.1 seconds for the \textit{baseline} policy and 22.3 seconds for the \textit{proposed} policy).
The error in the \textit{baseline} policy got larger, especially around the middle steps where different transition branches exist.
On the other hand, it can be seen that the \textit{proposed} policy effectively used the sensor data to forecast the timing of different target steps.

\subsection{Discussion}
\label{sec:study1-disc}

The results demonstrated the clear effectiveness of the proposed approach and its implications.
First, when tasks are complex and feature multiple potential transition pathways, employing stochastic models to predict user behavior constantly updated by sensor observation proves particularly beneficial.
It could be contended that the sequence of steps ought to remain constant, especially for routine tasks.
However, we chose to allow users the flexibility to adjust the order of steps as per their specific circumstances.
Such flexibility enables broader user scenarios as discussed in~\secref{sec:proposed-scenario}.
Simultaneously, enabling the system to learn from the same user's data over days represents a promising direction for future deployment.

The results also corroborated that uncertainty remains.
Tracking all steps in complex, everyday tasks with a common device like a smartwatch is quite challenging.
The limited tracking accuracy of certain target steps affects the behavior modeling around them.
Still, an effective intervention system does not always need to be built on highly accurate sensing. 
For example, step counters can be inaccurate sensors~\cite{thorup2017accuracy} but still an effective tool for behavior change~\cite{qiu2014step}.
Thus, we examine the utility of our intervention design in Study~2.

\section{Real-Time Agent System Implementation}

\begin{figure}[t]
    \centering
    \includegraphics[width=\linewidth]{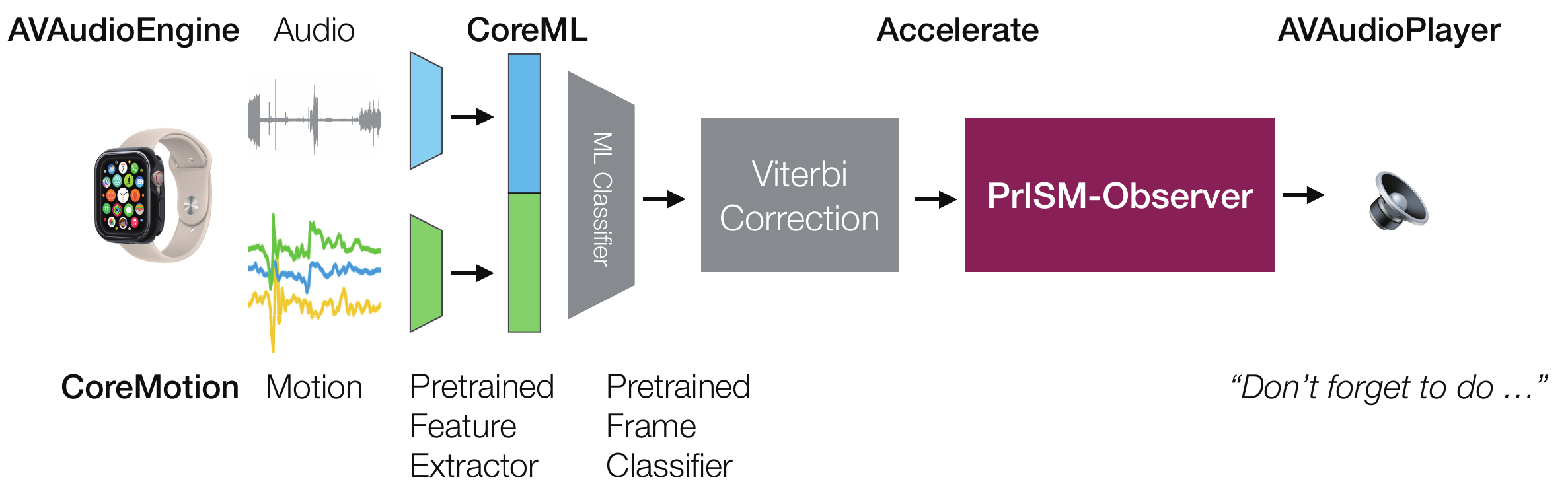}
    \caption{Implementation overview of the real-time agent system on Apple Watch.}
    \Description{This image is a flow diagram illustrating the architecture of a smartwatch-based system for task assistance. Starting on the left, data is collected through 'AVAudioEngine' for audio and 'CoreMotion' for motion, each producing waveform representations. These inputs feed into 'CoreML,' where a pre-trained feature extractor processes the data, which then passes through an ML classifier. The output from the classifier is subjected to 'Viterbi Correction' within the 'Accelerate' framework to enhance the prediction sequence. This processed data is finally analyzed by 'PrISM-Observer,' a monitoring module. The system finally uses the 'AVAudioPlayer' module, which provides auditory feedback to the user with reminders like 'Don't forget to do...'}
    \label{fig:implementation}
\end{figure}

Study~1 showed the proposed approach's effectiveness in forecasting the timing of target steps, based on which our interventions are triggered. 
To examine the utility of such real-time interventions, we developed a prototype using Apple Watch (Series 7, watchOS 10.3) with Swift.
Currently, there are two versions of the implementation: the laptop-server version and the watch-only version.
For the laptop-server version, we used a MacBook Pro with 16GB Apple M1 Chip as a server.
The Apple Watch streams the sensor data obtained through dedicated APIs (\ie AVAudioEngine and CoreMotion) via the network to the server in real-time.
The server runs PrISM-Tracker (frame-level HAR and the Viterbi correction), followed by PrISM-Observer (stochastic modeling and intervention policy).
When the intervention is to happen, the laptop plays the audio file.
This implementation is fast enough because it can utilize efficient computation libraries in Python.
We used this version for the user study (Study~2).

At the same time, we developed a watch-only version to show the feasibility of a self-contained agent on the device.
The process pipeline is shown in \figref{fig:implementation}.
Here, the obtained multimodal data is funneled into a pre-trained feature extractor that operates on the principle of CoreML~\cite{CoreML}, Apple's machine learning framework. 
The feature extractor analyzes the incoming data streams and translates them into a feature-rich format suitable for the subsequent frame-level classifier, which CoreML also powers.
Note that these models were already trained offline in the same manner with the prior work~\cite{DBLP:journals/imwut/ArakawaYMNRDRMC22} and converted into CoreML format beforehand using \texttt{coremltools}\footnotemark.
\footnotetext{\url{https://github.com/apple/coremltools}}
Following the classifier, the system applies the Viterbi tracking and intervention policy (including the stochastic modeling).
These components leverage Apple's Accelerate~\cite{Accelerate} for high-performance computations.
Finally, when the system decides to trigger interventions, the watch plays the corresponding audio file.
In both versions, note that while the system emits sound for intervention, the sensor readings of both motion and audio stop to prevent noise from adding to the tracking.

\begin{table}[t]
\caption{Average time taken to process one frame (equivalent to 0.2 seconds) at each module in our real-time agent system}
\Description{This table compares the performance time of different modules on a laptop versus a smartwatch. The table lists three modules: Frame-level HAR (Human Activity Recognition), Viterbi tracking, and Intervention policy. For each module, the time taken to execute on a laptop and a watch is given in seconds. The Frame-level HAR module takes 0.04 seconds on a laptop and 0.28 seconds on a watch. The Viterbi tracking takes considerably less time, with 0.0002 seconds on a laptop and 0.001 seconds on a watch. The Intervention policy takes 0.09 seconds on a laptop and 0.98 seconds on a watch.}
\label{tbl:speed}
\centering
    \begin{tabular}{r|c|c}
    \hline
    \textbf{Module}               & \textbf{Laptop (sec) } & \textbf{Watch (sec)} \\
    \hline
    Frame-level HAR     & 0.04 & 0.28        \\
    Viterbi tracking    & 0.0002  &  0.001        \\
    Intervention policy & 0.09  & 0.98   \\\hline
    \end{tabular}
\end{table}

We conducted a speed test by applying each module to a frame 500 times, the result of which is shown in~\tabref{tbl:speed}.
This result shows that the laptop-server version is fast enough to run the pipeline at 5~fps (Recall one frame is 0.2~seconds long).
However, the watch-only version is slower due to the current limitation in the computation power and lack of an efficient Swift library.
In detail, the most time-consuming computation is the Monte Carlo method to estimate the distribution of $D_t^{\hat{s}}$.
Thus, the current watch system runs PrISM-Tracker at 2~fps to update its internal state and PrISM-Observer once per 3 seconds to control intervention, which is still effective, provided each step has a certain duration.
We believe future engineering can further accelerate the watch-only version. 

\section{Study~2: User Study with a Real-Time Agent System in Cooking}

Finally, we conducted a user study to examine the accuracy of the real-time system and users' perception of the performance of in-situ interventions.
We used the laptop-server version implementation with the \textit{cooking} task introduced in Study~1, which showed that the task has relatively larger errors in forecasting the target step moment even though the proposed approach mitigates them.
This study involves evaluating the intervention effect where such uncertainty remains.

\subsection{System Configuration}

\begin{table*}[t]
\caption{Candidate intervention steps $\hat{S}$ in Study~2. Their intervention type was decided based on how accurately the step is detected by HAR and input by an external system designer. In actual cases, the selection is customizable based on user preferences.}
\Description{This table summarizes the performance of a system in recognizing specific task steps and the type of intervention applied based on the accuracy of the recognition. The table has four columns: Step, Error in Study 1 (sec), HAR Accuracy (F1), and Intervention Type. Each row details the listed steps s2, s6, s8, s11, and s14.}
\label{tbl:cooking-step-type}
\centering
    \begin{tabular}{r|c|c|c}
    \hline
    \textbf{Step}   & \textbf{Error in Study~1 (sec)} & \textbf{HAR Accuracy (F1)} & \textbf{Intervention Type} \\
    \hline
    $s_2$: washing hands with soap and water & 25.9 & 0.83  & \textsc{notify if forgotten}         \\
    $s_6$: pouring oil on the pan       & 72.3 & 0.64  & \textsc{remind in advance}        \\
    $s_8$: dropping a small amount of water & 85.4 & 0.28  &  \textsc{remind in advance}       \\
    $s_{11}$: wiping the pan surface  & 56.2 & 0.43  & \textsc{notify if forgotten}        \\
    $s_{14}$: cleaning the table   & 71.0 & 0.52  &  \textsc{notify if forgotten}       \\
    \hline
    \end{tabular}
\end{table*}

We first trained PrISM-Tracker~\cite{DBLP:journals/imwut/ArakawaYMNRDRMC22} (\ie frame classifier and transition graph $G$) as the tracking module and obtained the best thresholds ($h^{s_1}, h^{s_2}, ..., h^{s_{14}}$) for the intervention policy using all 17-session training data in the dataset.
Moreover, we decided the remaining parameters in the system, that is, $K^-$ and $K^+$, governing the timing of the \textsc{remind in advance} and \textsc{notify if forgotten} interventions around the target step, respectively.
The authors explored different values before the study and set $K^-=15$ and $K^+=15$ (seconds) as reasonable timing to trigger each type of intervention before and after the moment the target step is supposed to happen.

\begin{table*}
\caption{Participant information and their chosen steps for the intervention in Study~2.}
\Description{This table summarizes the participant's information such as gender, age, their frequency of doing the task. It also contains experimental information such as the experiment's environment, selected intervention steps, and intentionally skipped steps.}
\label{tbl:study2-cooking-participant}
\centering
    \begin{tabular}{r|c|c|c|c|c|c}
    \hline
    \textbf{Participant}& \textbf{Gender} & \textbf{Age} & \textbf{Frequency} & \textbf{Environment} & \textbf{Selected Steps} & \textbf{Skipped Steps}  \\
    \hline
    P1 & M & 20's & Occasional & kitchen 1 & $s_6$, $s_8$, $s_{11}$ & $s_{14}$\\
    P2 & M & 20's & Often & kitchen 1 & $s_2$, $s_6$, $s_8$, $s_{11}$, $s_{14}$ &  \\
    P3 & M & 20's & Often & kitchen 1 & $s_8$, $s_{11}$, $s_{14}$ & $s_{11}$\\ 
    P4 & F & 20's & Often & kitchen 1 & $s_2$, $s_6$, $s_8$, $s_{11}$, $s_{14}$ & $s_2$, $s_{14}$ \\
    P5 & F  & 20's & Very often & kitchen 1 & $s_2$, $s_{11}$, $s_{14}$ & $s_{11}$, $s_{14}$ \\
    P6 & F & 40's & Often & kitchen 2 & $s_2$, $s_6$, $s_8$, $s_{11}$ & $s_2$, $s_{11}$ \\
    P7 & M & 70's & Very often & kitchen 2 & $s_2$, $s_8$, $s_{14}$ & $s_2$, $s_{14}$ \\
    P8 & F & 60's & Very often & kitchen 2 &  $s_2$, $s_6$, $s_8$, $s_{11}$, $s_{14}$ & $s_{11}$\\
    P9 & M & 30's & Occasional & kitchen 2 &  $s_2$, $s_6$, $s_8$, $s_{11}$, $s_{14}$ & $s_2$, $s_{11}$, $s_{14}$ \\
    P10 & M & 20's & Occasional & kitchen 2 & $s_8$, $s_{11}$, $s_{14}$  & $s_{14}$\\\hline
    \end{tabular}
\end{table*}

In addition, we chose candidate steps $\hat{S}$ from all the 14 steps that users may forget or want to be reminded of (Recall the step selection process shown in \figref{fig:step-selection}).
Here, two system designers individually chose candidate steps first and then had a discussion to reach an agreement.
As a result, they chose five steps: $s_2$ washing hands with soap and water, $s_6$ pouring oil on the pan, $s_8$ dropping a small amount of water, $s_{11}$ wiping the pan surface, and $s_{14}$ cleaning the table. 
Moreover, as described in~\secref{sec:proposed-formulation}, the system suggests the intervention type option based on each step's detectability by frame-by-frame HAR (without the Viterbi correction~\cite{DBLP:journals/imwut/ArakawaYMNRDRMC22}, corresponding to the left figure in~\figref{fig:cooking-raw-result} in~\appref{app:prism-tracker-performance}).
The F1-score for the chosen steps were $0.83$, $0.64$, $0.28$, $0.43$, and $0.52$, respectively.
We first decided to assign \textsc{remind in advance} to $s_{11}$ due to its low accuracy.
Moreover, the system designers suggested that doing $s_6$ before putting an item on the pan is important in actual cooking, and thus, we assigned \textsc{remind in advance} to $s_6$ as well.
We assigned \textsc{notify if forgotten} to the other steps.
Note that we used the same intervention set for all participants in this study to make the comparison meaningful, but the end-users would always have the flexibility to switch the type.
The detection accuracy and the assigned intervention type for each $\hat{s} \in \hat{S}$ are summarized in~\tabref{tbl:cooking-step-type}.

\subsection{Procedure}

We recruited 10 participants (P1 -- P10) through word-of-mouth. 
They self-reported how often they cook in their daily lives and were all right-handed.
To test the robustness against different environments, we used the same kitchen (kitchen 1) as in the training data for five participants and a different kitchen (kitchen 2) for the rest of the participants. 
Their demographic information is presented in~\tabref{tbl:study2-cooking-participant}.

\subsubsection{Step Selection}

After consent, we showed a list of all the steps $S$ and a tutorial video.
Then, we shared interaction candidate steps $\hat{S}$, and asked which steps the users believe they might miss at times and would prefer some intervention support.
We asked them to choose at least two steps for the experiment.
We also encouraged them to imagine the task would be their routine and which intervention they might want.
Simultaneously, we emphasized that they could change the step order flexibly without having to follow the tutorial.
We configured the agent system once they finalized the steps.

Here, if their chosen steps included a step for the intervention of \textsc{notify if forgotten}, we asked them to intentionally skip some of them to test the accuracy of the intervention policy.
Note, if the participant intentionally skipped the last step $s_{14}$, the study ran for an extra $K^+$ + 23.6 (the average time for $s_{14}$) seconds to give the system enough time to judge whether to trigger the \textsc{notify if forgotten} intervention for $s_{14}$.
We did not specify what to do during the extra time, and the participants behaved naturally.
The selected steps for each participant and which step they skipped are also shown in~\tabref{tbl:study2-cooking-participant}.

\subsubsection{Task Execution}

Once we configured the system, we answered their questions about the task to ensure they understood the procedure.
We also explained that, during the task execution, they were expected not to pause for questions (except in case of an emergency).
They pressed the button on the watch app to begin the task and hit the button again to end the task.
After the task, they completed a questionnaire, followed by semi-structured interviews about their experience of the system.
The entire session took approximately 30 minutes for one participant.

\subsection{Metric}

We measured the timing difference between the triggered intervention and the target step.
Here, the timing for the target step was the moment the participant finished the previous step, which is the same timing as they started the target step unless they intentionally skipped the target step.
Note that the triggered intervention timing here accounted for the offset, that is, $K^+$ and $K^-$.
We manually annotated these timings for each session.

In addition, for the \textsc{notify if forgotten} intervention, we annotated if each intervention was accurate. True positive (TP) when it was triggered and the participant skipped the target step; false positive (FP) when it was triggered and the participant did the step; false negative (FN) when it was not triggered and the participant skipped the step; and, true negative (TN) when it was not triggered and the participant did the step.

In the post-task questionnaire, we asked the participants to rate the timing accuracy of each intervention they experienced.
Note, for the \textsc{notify if forgotten} intervention, we asked about the timing accuracy only if it was triggered (\ie either TP or FP).
They answered whether each intervention was accurate in terms of timing with a Likert scale from 1 (strongly disagree) to 7 (strongly agree).
Then, they answered whether the system is reliable with a Likert scale from 1 (strongly disagree) to 7 (strongly agree).
In addition, we asked for their behavioral intention, guided by the Technology Acceptance Model~\cite{Davis1989Perceived}, which explains users' attitudes towards technologies and is frequently used to evaluate how likely individuals are to use the technologies.

\subsection{Results}

\begin{figure}[t]
    \centering
    \includegraphics[width=\linewidth]{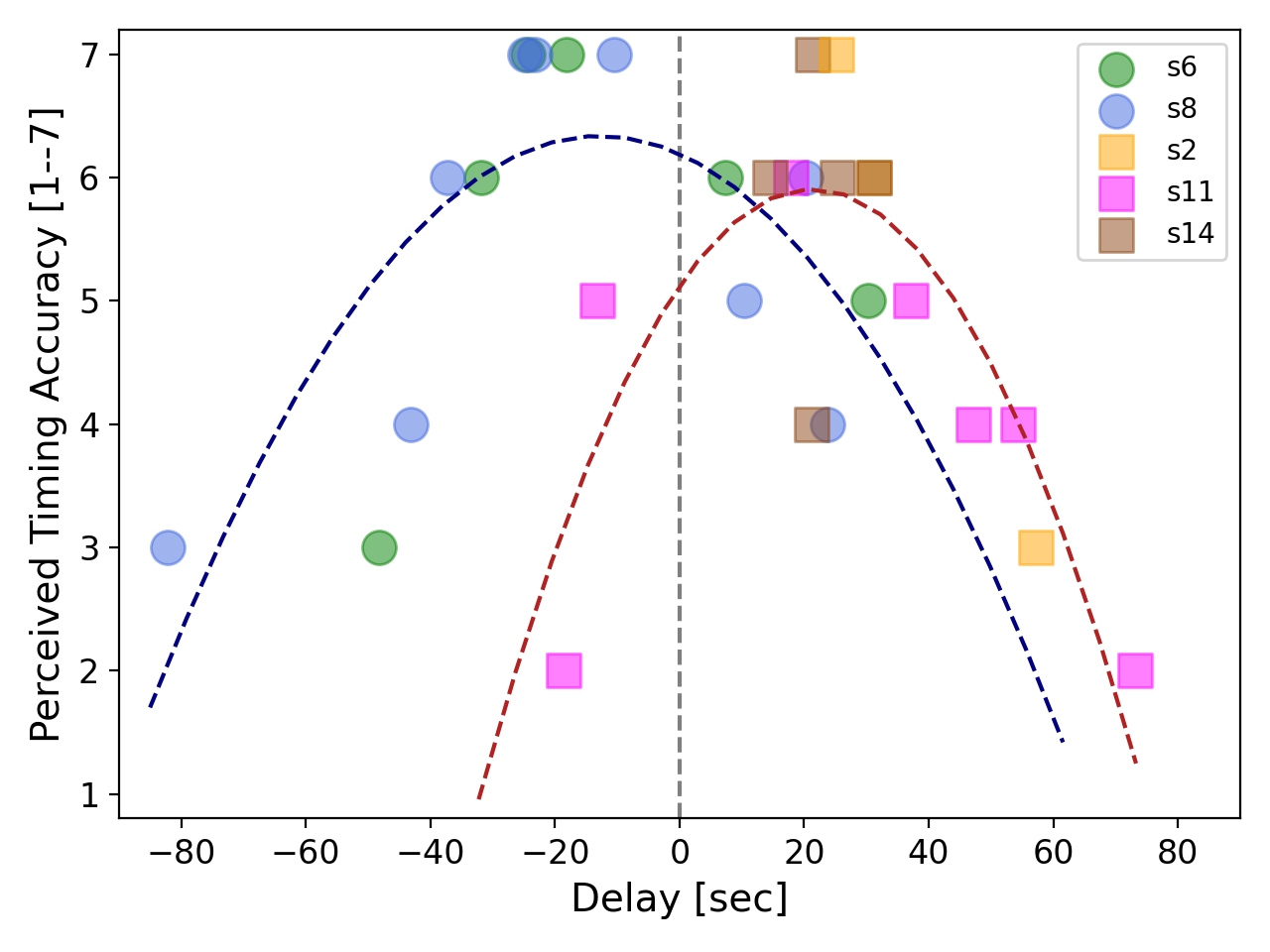}
    \caption{The delay of each triggered intervention from the actual timing of the target step and its perceived timing accuracy. The shape indicates the intervention type (circle: \textsc{remind in advance} and square: \textsc{notify if forgotten}). The color indicates different steps. The two dashed lines are trend lines obtained by fitting a quadratic function separately to each intervention type (blue: \textsc{remind in advance} and red: \textsc{notify if forgotten}). }
    \Description{This scatter plot with fitted lines illustrates the perceived timing accuracy of interventions at various delays for different task steps. The x-axis represents delay in seconds, ranging from -80 to +80 seconds, and the y-axis measures the perceived timing accuracy on a scale from 1 to 7. Colored data points represent different steps in a task, with green for s6 (circle), light blue for s8 (circle), pink for s2 (square), brown for s11 (square), and orange for s14 (square). The plotted points are spread across two bell-shaped curves: one is dashed blue (for circle) and the other dashed red (for square), indicating different intervention types.}
    \label{fig:study2-delay-perception}
\end{figure}

\figref{fig:study2-delay-perception} shows the delay of each intervention and how accurate the participants felt about the timing of the intervention.
The circle markers indicate the \textsc{remind in advance} intervention (\ie $s_6$ and $s_8$) and the square markers indicate the \textsc{notify if forgotten} intervention (\ie $s_2$, $s_{11}$, and $s_{14}$).
The result suggests that the agent system could trigger most interventions with a small timing error, and the participants felt they were accurate (20 out of the total 27 triggered interventions were rated five or higher on the 1--7 Likert scale).
Also, by looking at the trend lines, the relationship between the delay and perceived accuracy is implied. 
For instance, the participants perceived the \textsc{remind in advance} intervention (circle markers) as inaccurate as the delay becomes positively larger. At the same time, it was best rated when triggered some seconds before the delay is 0.
This is aligned with the type of the intervention---users want it to be triggered before the target step---validating our design, specifically, the use of the offset $K^-$ with $E[D_t^{\hat{s}}]$.
Conversely, they rated higher for the \textsc{notify if forgotten} intervention (square markers) if they happen some seconds after, also corroborating the use of the offset $K^+$.
While more samples are demanded to examine the trend, we can infer that the participants' expectations about the different interventions affected their perception of the intervention timing.

Additionally, there was no significant difference in the delay between kitchens 1 and 2 ($p > 0.05$) in this cooking scenario.
While more investigation is demanded to quantify the effect, this result suggests the benefit of the smartwatch-based agent system: the location-independent HAR capability.

\begin{figure}[t]
    \centering
    \includegraphics[width=0.75\linewidth]{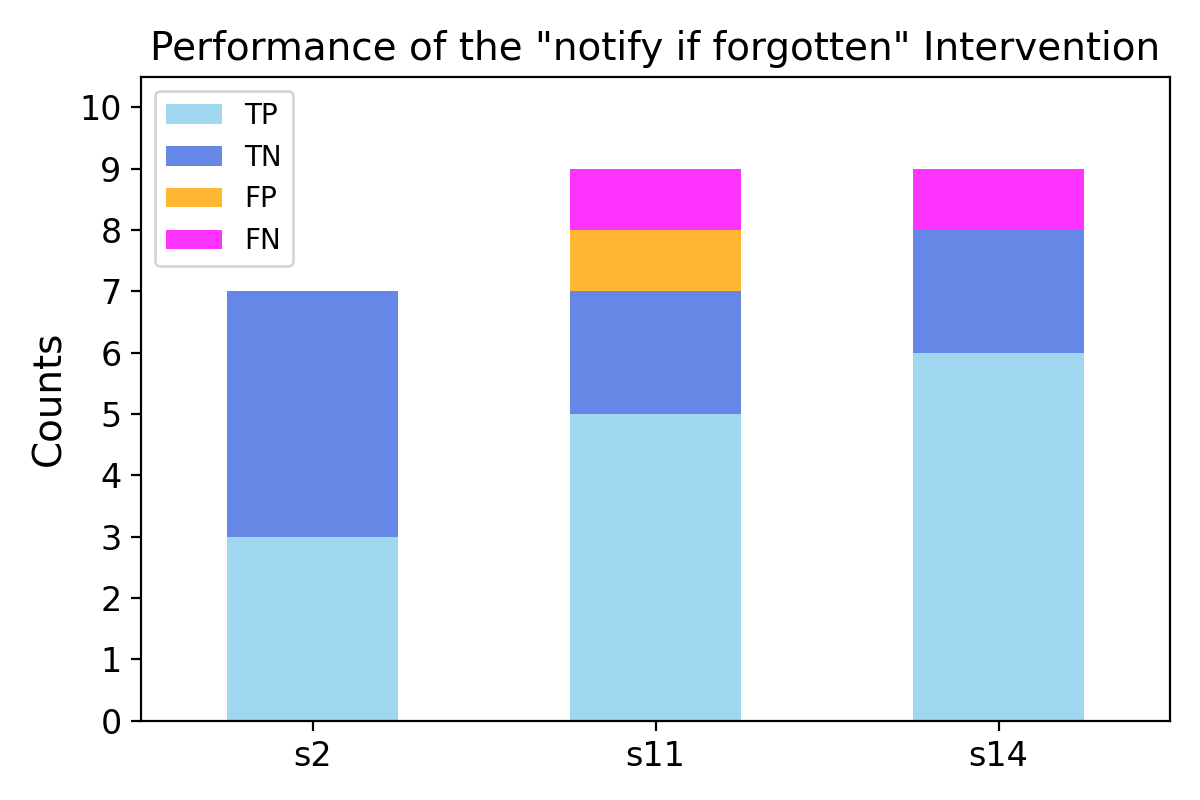}
    \caption{Performance of each of the \textsc{notify if forgotten} interventions. TP, TN, FP, and FN indicate true positive, true negative, false positive, and false negative, respectively.}
    \Description{This bar chart represents the accuracy of the 'notify if forgotten' intervention based on counts of true positives (TP), true negatives (TN), false positives (FP), and false negatives (FN) for three task steps: s2, s11, and s14. Each bar is divided into segments that correspond to the counts for each accuracy category, with light blue for TP, dark blue for TN, orange for FP, and pink for FN.}
    \label{fig:study2-notification-accuracy}
\end{figure}

\figref{fig:study2-notification-accuracy} shows whether each of the \textsc{notify if forgotten} interventions was accurate.
While the number of samples is small, the results of the low error rate (3 out of 25 chances) indicate that the agent system could detect whether the target step happened or not and notify the participants appropriately.
More specifically, there was no false intervention for $s_2$ thanks to the high HAR accuracy (See \tabref{tbl:cooking-step-type}).
On the other hand, there were false interventions for $s_{11}$ and $s_{14}$, which can be explained by a relatively lower HAR accuracy.
Given this, the agent system could infer the risk of false interventions in the model training phase, which would help system designers configure the system.

\begin{figure}[t]
    \centering
    \includegraphics[width=0.75\linewidth]{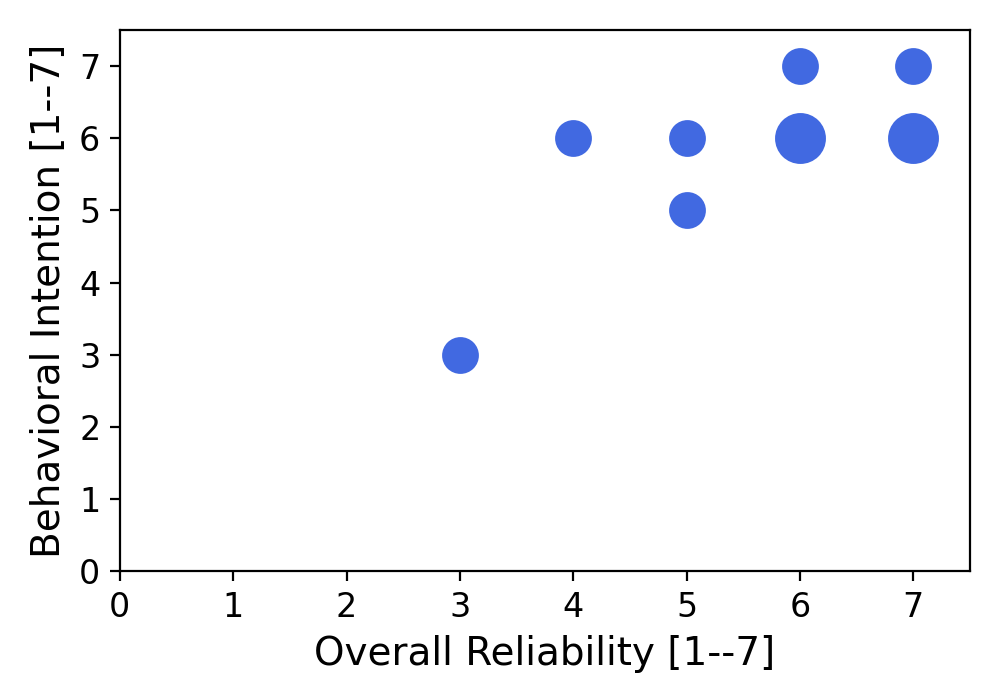}
    \caption{The rated reliability and behavioral intention of the agent system. Each dot represents a participant's answer to the questionnaire. The larger circles indicate two participants.}
    \Description{This scatter plot visualizes the relationship between overall reliability and behavioral intention on a scale of 1 to 7. The x-axis indicates 'Overall Reliability,' and the y-axis represents 'Behavioral Intention,' both measured on a Likert scale from 1 (low) to 7 (high). The data points, all in shades of blue, appear to show no clear trend or correlation, with one point at the lower end of both scales around (3,3) and the remaining points scattered across the higher end of the scales, particularly with behavioral intention scores of 6 and above.}
    \label{fig:study2-reliability-intention}
\end{figure}

\figref{fig:study2-reliability-intention} presents the relationship between the overall reliability and the behavioral intention answered by the participants.
While a longer-term study is ideal, this shows initial evidence that such a task-support agent system is favorably accepted.
The result also suggests a correlation between reliability and behavioral intention, which implies that reducing errors and achieving situated interventions is important.

\subsection{User Comments}

Finally, we summarized the qualitative insights from the semi-structured interviews after the task.

\subsubsection{How Participants Selected Interventions}

The number of participants who enabled each target step $\hat{s} \in \hat{S}$ in the step selection phase is 7, 6, 8, 9, and 5, for $s_2$, $s_6$, $s_8$, $s_{11}$, and $s_{14}$, respectively.
The individual difference shown in \tabref{tbl:study2-cooking-participant} reflects their preference.
For instance, P7 mentioned, ``I never forget pouring oil when using the pan, so not including $s_2$.''
P3 said, ``I thought I would be inattentive in the latter part of the procedure.''
P5 said, ``I'm experienced. I only need them if I make a mistake.''
P6 commented, ``I am a careless person. I would enable all notifications. I believe I would not be annoyed very much. It'd be like a fun assistant.''
In addition to these comments, they agreed that using different types of intervention made sense.
P8 said, ``I don't need to be reminded every time to wash hands, so the system notifying me only if I make a mistake is a good design.''
The results underpin the design of our framework, enabling end-user customization of the intervention.

\subsubsection{How Participants Perceived Interventions}

Next, we asked about the interventions they remembered most and their impression of the interventions.
They were surprised by the agent's capability, especially for the \textsc{notify if forgotten} intervention.
P1 said, ``I'm amazed by how it knows about my situation. The notification (for $s_{14}$) was accurate.''
At the same time, a few participants mentioned a need for certain notifications to be more precise.
P9 mentioned, ``For $s_{11}$ (wiping the pan surface), it was after I already put my sausage, and thus I felt it is a bit inaccurate.''
P3 also suggested that it is safe to trigger interventions earlier as the steps can be irreversible in the cooking.
Our framework is flexible enough to incorporate such adjustments, for example, by using a smaller $K^{+}$ value to enable immediate detection or changing the type of the intervention into \textsc{remind in advance}.

Similarly, regarding the \textsc{notify if forgotten} intervention, P2, after experiencing FP (false positive), said, ``I was not annoyed much by the notification ($s_{11}$). I just thought it was checking me.''
On the other hand, P1 experienced FN (false negative), mentioning, ``I was a bit sad when it did not happen even though I skipped it ($s_{14}$). In the real world, strict checking may be good for me.''
The results suggest the tendency of false-positive resistance and false-negative sensitiveness, consistent with a design strategy recognized in research on real-time interventions that rely on imperfect sensing~\cite{DBLP:conf/chi/ArakawaY21}.

Moreover, the participants agreed that the \textsc{remind in advance} intervention is useful for unfamiliar steps like $s_{8}$.
They also suggested that it should happen before the step. 
P9 mentioned, ``Earlier is good. If it is after I have done the step, it would sound hilarious. But too early is also not good. Being able to feel some intelligence in the system is what attracts me.''
This comment highlights the importance of minimizing timing errors and the design of interventions that meet user expectations.

\subsubsection{Further Use Cases}

Acknowledging the capability of the agent system, the participants shared an interest in extending the application.
The examples include skin care (3 participants), gym/exercise routine (3 participants), furniture assembly (2 participants), medication (2 participants), house cleaning (2 participants), using a laundry room (1 participant), and using a maker space (1 participant).
For example, P1 mentioned, ``I sometimes forget to sanitize the machine after I use it in the gym. Also, it would help me complete my exercise routine without forgetting an activity.''
Supporting these various needs is promising, given \proposed{}'s generalizable performance suggested in Study~1.

\subsubsection{Room for Improvement}
\label{sec:study2-comments-room}

Finally, we also gained insights about future improvement.
P8 said, ``The system did not work well for me throughout the task. It was wrong from the beginning. I wanted to fix it by telling it.''
For this user, the system mistook the path P8 followed; P8 cooked the sausage first while the system guessed she cooked an egg first due to HAR error, resulting in overly early interventions.
In such a case, as P8 expressed, offering a way for the user to correct the agent's belief will be ideal, which we discuss further later in~\secref{sec:disc-interactivity}.

On another note, P1 mentioned, ``I would like to see more variations in the audio message with different voice styles. If it were my favorite idol's voice, I might accept it even if the intervention is wrong.''
Such a social aspect between the task-support agent system and the user is an exciting research area~\cite{DBLP:journals/csur/SeabornMPO21}.

\subsection{Summary}

The results demonstrated the accuracy of our agent system's intervention, which the participants favorably accepted.
The user comments highlighted the benefits of \proposed{}'s design, such as the flexibility to cater to different user needs and to adjust parameters to enhance the user experience.
We believe Study~2 produces insights into the human-AI interaction system powered by sensing technology in the physical world.

It should be noted that although seven participants completed the task in a sequence differing from that presented in the tutorial, we cannot rule out the possibility of a study-induced bias affecting task behavior. Acknowledging this limitation, we plan to undertake a longer-term study to assess the system's ability to adapt to spontaneous user behavior during tasks.

\section{Limitations and Future Work}

The current formulation and implementation of \proposed{} is not without limitations.
We also describe future work to advance the field further.

\subsection{Long-Term Behavior Study}

Study~2 is in a controlled setting and thus does not perfectly reflect the user behavior performing routine tasks under natural conditions, such as inattentiveness.
A long-term real-world study is needed to further investigate the system's benefits and user perception. 
We want to emphasize that user behavior should be considered flexible instead of a fixed sequence for reasonable human-AI interaction so the system can adjust to spontaneous user behavior, such as switching the order of the steps.
We believe \proposed{}'s capability to incorporate such uncertainty in human behavior will be a fundamental solution.

In addition, a long-term study will also shed light on the agent's capability to learn from user behavior to adapt the intervention policy. 
For instance, if the user often forgets a certain step, the system switches the intervention from \textsc{notify-if-forgotten} to \textsc{remind-in-advance}.
Conversely, if the user gets accustomed to the task enough and thinks \textsc{remind-in-advance} is too much, it can be switched to \textsc{notify-if-forgotten}.
It is of interest to examine the optimal balance of the agent's adaptability and the end-user's controllability.

\subsection{Health and Accessibility Applications}


The proposed framework will be especially helpful for health applications to assist people in need.
For example, people with dementia struggle to perform everyday routine tasks properly, and thus, interventions are helpful~\cite{Josephsson1993Supporting, Lazarou2016Novel}.
Moreover, patients who need to perform self-care after surgery regularly could benefit from it to avoid the critical consequence of infection due to mistakes~\cite{shandra2016health, levin1983self}.
We are currently developing and evaluating assistants for postoperative wound care with skin-cancer patients undergoing Mohs micrographic surgery~\cite{Vaccarello2024}.

\subsection{Detecting Errors within a Step}

\proposed{} can detect and remind a user of a step in the procedure.
Currently, it does not support dealing with more fine-grained errors, such as errors within a step, often necessitating visual information to be detected.
For example, while steaming milk for the latte-making task, it is important to keep the jug at a proper angle to get a good milk texture, which the current system cannot sense.
While aligning user expectations about the system's capability is crucial, we plan to investigate further sensing capability, such as applying multimodal anomaly detection~\cite{DBLP:conf/chi/ArakawaY19} to compare the sensor data within the step with prior ``good'' behavior.

\subsection{Refining Step Granularity}

Automating and optimizing the process of dividing the procedure into steps is crucial to scaling our framework to diverse tasks.
In this regard, Zhou~\etal~\cite{DBLP:conf/acl/ZhouZY0YCN22} proposed an approach to constructing an open-domain hierarchical knowledge base of procedures.
Wake~\etal~\cite{DBLP:conf/sii/WakeAYKSTI21} created a task model for decomposing human demonstration of procedural tasks for robots to model the process.
In our case, it could be possible for the framework to use an automatically generated transition graph internally to model the user behavior and to trigger intervention.
In contrast, the user specifies the desired intervention without being aware of the representation of the steps used by the system. 

\subsection{Interweaving Broader Interactions}
\label{sec:disc-interactivity}

This work focuses on passive interaction from the user's perspective; the system monitors user task behavior and proactively triggers intervention.
Though our study demonstrated its effectiveness, the potential role of a task-support intelligent agent can be more versatile~\cite{DBLP:conf/chi/JaberZKHGBCM24}.
For instance, more dialogue between the user and the system could be an interesting area to study.
PrISM-Tracker~\cite{DBLP:journals/imwut/ArakawaYMNRDRMC22} showed a preliminary approach to update its tracking belief through dialogue.
In the framework of \proposed{}, if the system could ask the user ``What are you doing?'', the system could resolve the uncertainty about the current step.
Similarly, asking ``What will you do next?'' would resolve the uncertainty about the future step transition. 
Conversely, the user's reaction to the intervention could correct errors in the agent, as suggested in~\secref{sec:study2-comments-room}.
Likewise, if there is a question-answering capability, the user's asking, ``What should I do after washing my hands?'' would also help the agent track the user's state.
Given the surge of human-like chat capability enabled by large language models, investigating such real-time dialogue interactions along with sensor data is promising.

\subsection{Extending to Other Sensing Platforms}

We implemented our prototype on a smartwatch based on its capability to sense a user across various daily activities at different places. 
At the same time, the framework's stochastic modeling can be used as a post-process for other HAR systems.
For example, \textit{VAX}~\cite{DBLP:journals/imwut/PatidarGA23} uses ambient, privacy-sensitive sensors such as Doppler RADARs and LIDARs. 
We can install such a HAR system in the user's kitchen to monitor their everyday cooking and use \proposed{} on top of it.
Existing voice assistants like Alexa could also be a platform for integrating our framework, leveraging acoustic sensing to obtain user context.
In the context of voice assistant, while prior research~\cite{frummet2024cooking, Chu2021Recipe} has shown the benefit of dialogue-based guidance of steps, our work extends their capability by context awareness and user-centered intervention design, as recently emphasized by Jaber~\etal~\cite{DBLP:conf/chi/JaberZKHGBCM24}.
Investigating different input sensors and their effect on end-user experience is an important future study.

\section{Conclusion}

We presented \textit{\proposed{}}, a framework for designing interventions to mitigate errors in daily procedural tasks (\eg forgetting a step), and developed a real-time agent system on a smartwatch leveraging multimodal Human Activity Recognition (HAR).
The stochastic modeling of user behavior and intervention policy based on it enables situated triggering of interventions.
Moreover, the framework is designed so users or system designers can customize which step they need intervention for and how.
Study~1, involving three daily task datasets, verified the proposed approach's effectiveness in optimizing the intervention timing.
In addition, Study~2, using the real-time smartwatch system in the cooking scenario, resulted in positive participant feedback, providing qualitative insights into such real-time intervention in procedural tasks.
The results and discussion pave the way for achieving reliable human-agent interaction where the agent's sensing capability is limited, which is often the case in real-world tasks.

\bibliographystyle{ACM-Reference-Format}
\bibliography{paper}

\appendix

\section{PrISM-Tracker's Performance}
\label{app:prism-tracker-performance}

\begin{figure*}[h]
    \centering
    \includegraphics[width=0.75\linewidth]{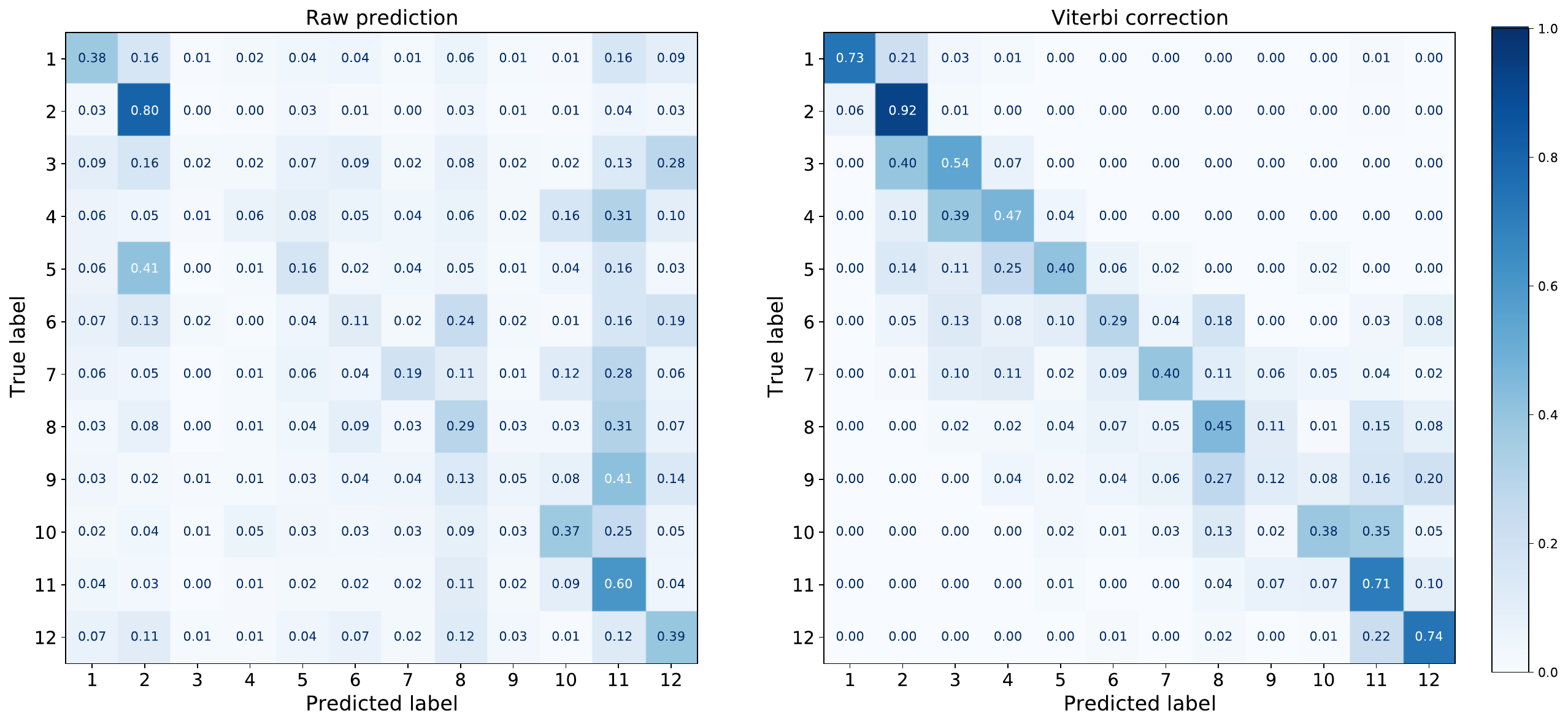}
    \caption{Frame-level confusion matrix on the wound care task. (left) raw HAR (right) after the Viterbi correction.}
    \Description{This image showcases two confusion matrices side by side, comparing raw prediction and Viterbi correction outcomes for a classification task with 12 categories for the wound-care task. Each matrix has both predicted labels (x-axis) and true labels (y-axis) ranging from 1 to 12. The color intensity represents the frequency of predictions, with darker shades of blue indicating higher frequencies.}
    \label{fig:wound-care-raw-result}
\end{figure*}

\begin{figure*}[h]
    \centering
    \includegraphics[width=0.75\linewidth]{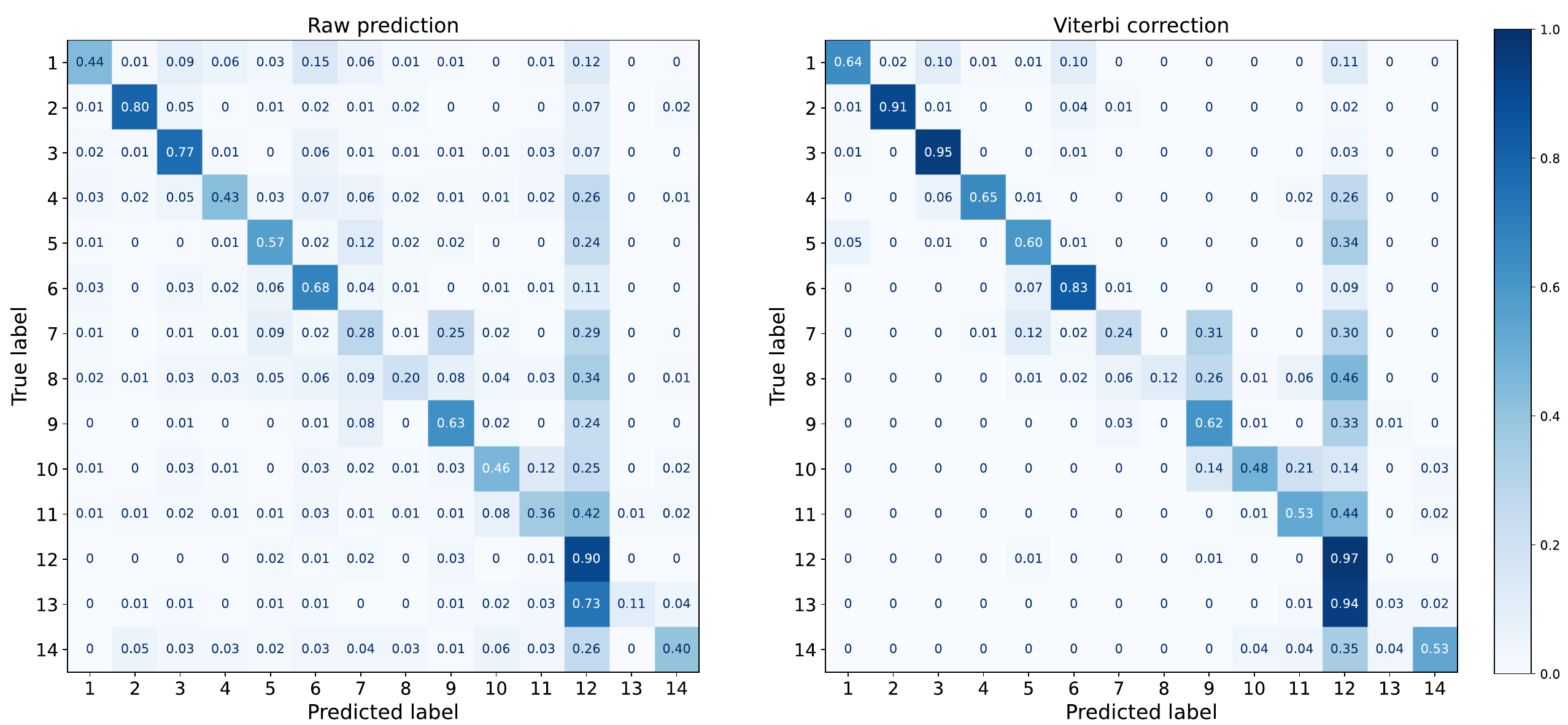}
    \caption{Frame-level confusion matrix on the cooking task. (left) raw HAR (right) after the Viterbi correction.}
    \Description{This image showcases two confusion matrices side by side, comparing raw prediction and Viterbi correction outcomes for a classification task with 14 categories for the cooking task. Each matrix has both predicted labels (x-axis) and true labels (y-axis) ranging from 1 to 14. The color intensity represents the frequency of predictions, with darker shades of blue indicating higher frequencies.}
    \label{fig:cooking-raw-result}
\end{figure*}

\begin{figure*}[h]
    \centering
    \includegraphics[width=0.75\linewidth]{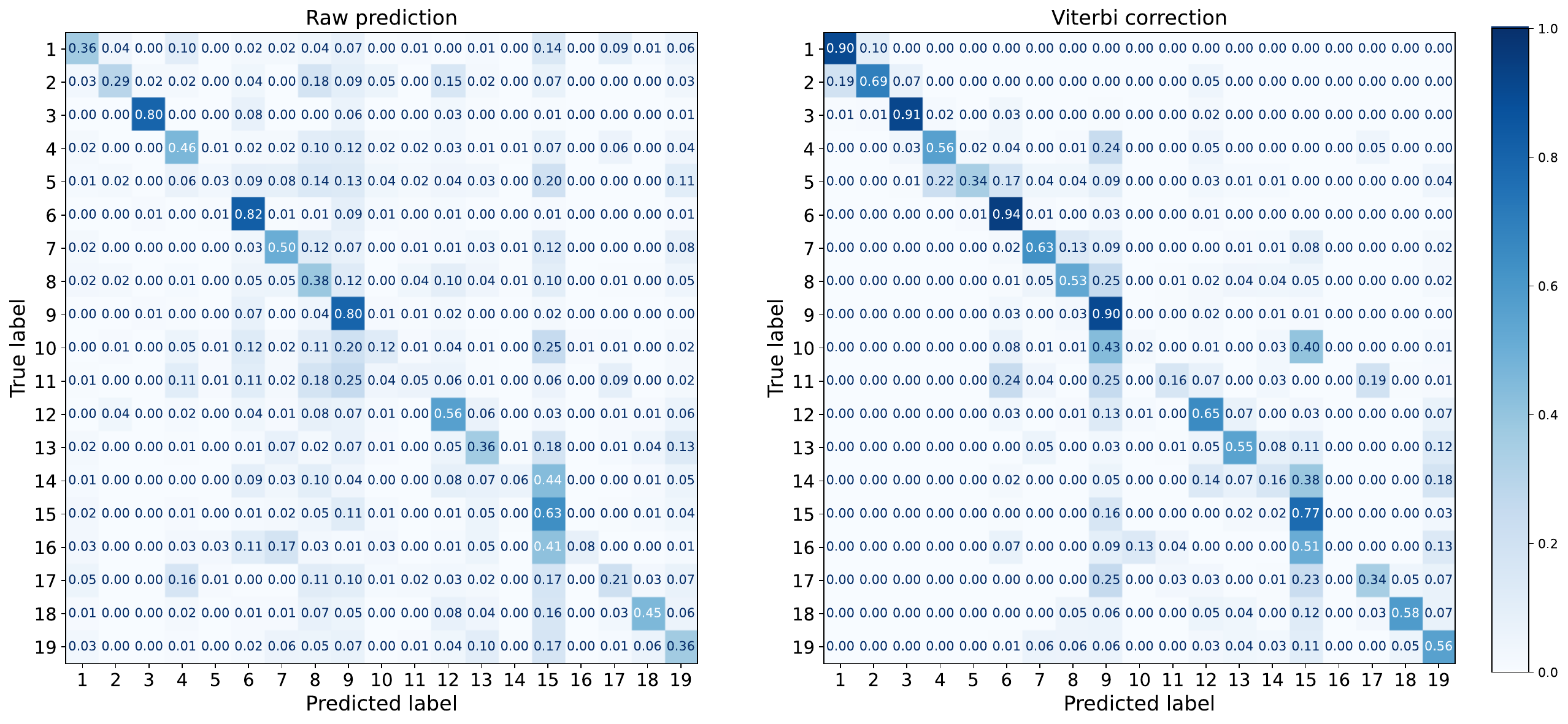}
    \caption{Frame-level confusion matrix on the latte-making task. (left) raw HAR (right) after the Viterbi correction.}
    \Description{This image showcases two confusion matrices side by side, comparing raw prediction and Viterbi correction outcomes for a classification task with 19 categories for the latte-making task. Each matrix has both predicted labels (x-axis) and true labels (y-axis) ranging from 1 to 19. The color intensity represents the frequency of predictions, with darker shades of blue indicating higher frequencies.}
    \label{fig:latte-making-raw-result}
\end{figure*}

\figref{fig:wound-care-raw-result} -- \figref{fig:latte-making-raw-result} show the frame-level HAR confusion matrix of PrISM-Tracker~\cite{DBLP:journals/imwut/ArakawaYMNRDRMC22} without and with the Viterbi correction, for the wound care, cooking, and latte-making task, respectively.
Recall that one frame corresponds to 0.2 seconds.
While the Viterbi correction improved the classification performance, inaccuracy remains, especially for steps with similar signal profiles: macro F1-Scores for the wound care, cooking, and latte-making tasks were 50.7\%, 61.5\%, and 52.9\%, respectively.

\section{Details of Intervention Policy Implementation}
\label{app:policy-implementation-detail}

The intervention policy described in~\secref{sec:proposed-policy} includes several hyperparameters.
First, there is a step-dependent entropy threshold $h^{\hat{s}}$, which governs the timing to initiate an intervention timer.
Then, after the intervention timer starts, the framework keeps monitoring $E[D_t^{\hat{s}}]$. If there is a significant change within the next $p = 10$ seconds (before the timer ends), the timer is discarded, for which we use a threshold parameter $e = 30$ seconds.
Moreover, as shown in~\figref{fig:timer}~Bottom, the entropy fluctuates due to the randomness in the Monte-Calro method, thereby necessitating the need for smoothing.
We apply a moving average smoothing with the size of $w = 2$ seconds to the entropy.
Parameters $p$, $e$, and $w$ were empirically determined in the initial observation of $D_t^{\hat{s}}$ transitions of a few sessions.
On the other hand, the entropy threshold $h^{\hat{s}}$ was optimized through grid search in the process of the leave-one-session-out cross-validation.

In addition, for the \textsc{notify if forgotten} intervention in the real-time system used for Study~2 (cooking), the system monitors the user action from the moment the timer started for $E[D_t^{\hat{s}}] + K^+$ seconds to judge whether the target step $\hat{s}$ happens. 
Since frame-level prediction can fluctuate due to sensing noise, we apply a moving average smoothing with the size of 1 second, which corresponds to 5 frames.
Moreover, the system judges the step $\hat{s}$ happens if the probability for the step is highest for 5 seconds, which was also determined empirically based on the fact that every step in the cooking usually lasts more than 10 seconds (See~\figref{fig:dataset-graph}).


\end{document}